\documentstyle[11pt]{article}

\font\srm = cmr9

 \font\fmi=cmitt10 scaled \magstep2

\newcommand{\Title}[1]{\title{\bf #1}}

\begin{document}

\newcommand{\be}{\begin{equation}} 
\newcommand{\fe}{\end{equation}}
\newcommand{\eqn}{\label}
\newcommand{\bel}{\begin{equation}\label}

\def\thf{\baselineskip=\normalbaselineskip\multiply\baselineskip
by 7\divide\baselineskip by 6}


\def\fff{\baselineskip=\normalbaselineskip}


\def\spose#1{\hbox to 0pt{#1\hss}}\def\lta{\mathrel{\spose{\lower 3pt\hbox
{$\mathchar"218$}}\raise 2.0pt\hbox{$\mathchar"13C$}}}  \def\gta{\mathrel
{\spose{\lower 3pt\hbox{$\mathchar"218$}}\raise 2.0pt\hbox{$\mathchar"13E$}}} 

\def\Libra{\spose {--} {\cal L}}
\def\Diam{\spose {\raise 0.3pt\hbox{+}} {\diamondsuit}  }

\font\fiverm=cmr5
\def\d{\delta}
\def\dL{\spose {\lower 5.0pt\hbox{\fiverm L} } {\delta}}
\def\dE{\spose {\lower 5.0pt\hbox{\fiverm E} } {\delta}}
\def\DL{\spose {\lower 5.0pt\hbox{\, \fiverm L} } {\Delta}}
\def\DE{\spose {\lower 5.0pt\hbox{\, \fiverm E} } {\Delta}}

\def\eqdef{\fff\ \vbox{\hbox{$_{_{\rm def}}$} \hbox{$=$} }\ \thf }

\def\ov{\overline}  

\def\uth{{\,^{\,_{(3)}}}\!}  \def\utw{{\,^{\,_{(2)}}}\!}
\def\ud{{\,^{\,_{(\rm d)}}}\!}  \def\udi{\,{^{\,_{(\rm d-1)}}}\!}
\def\up{{\,^{\,_{(\rm p)}}}\!}  \def\udp{{\,^{\,_{(\rm d)}}}\!}

\def\Df{{\cal D}}    \def\Ff{{\cal F}}
\def\Kf{{\cal H}}    \def\Xf{{\cal X}}
\def\Af{{\cal A}}    \def\af{{\alpha}} 
\def\Qf{{\cal Q}}    \def\Pf{{\cal P}} 
\def\kil{\hbox{\fmi k}}          \def\el{\ell}     \def\rad{{\hbox{\fmi r}}}

\def\bg{{\,\hbox{\fmi g}\,}} \def\nabl{{\nabla\!}}
\def\bepsilon{{\varepsilon}}
\def\calR{{{\cal R}}} \def\calW{{{\cal W}}}
\def\calS{{{\cal S}}}

\def\bPhi{{\mit\Phi}} \def\bF{{\, \hbox{\fmi F}\,}}
\def\bN{{\,\hbox{\fmi F}^{_{\,\{\rm r+1\}}}\!}}   \def\bD{{\,\hbox{\fmi D}\,}}

\def\bA{{\,\hbox{\fmi A}\,}} \def\bB{{\,\hbox{\fmi A}^{_{\{\rm r\}}}\!}}

\def\oD{{\ov{\,\hbox{\fmi D}\,}}}
\def\hlag{{\hat{\cal L}}} \def\hT{{\hat{\,\hbox{\fmi T}\,}}}
\def\hj{{\hat{\,\hbox{\fmi j}\,}}}  
\def\hW{{\hat{\,\hbox{\fmi j}_{\!_{\{\rm r\}}}\!}}} 
\def\hf{{\hat{\,\hbox{\fmi f}}}}
\def\vv{{\,\hbox{\fmi v}\,}}     \def\ie{{\,\hbox{\fmi e}\,}}
\def\oP{{\ov{\,\hbox{\fmi P}\,}}} \def\calP{{\cal P}}
\def\caloP{{\ov{\cal P}}}
\def\Xx{{\,\hbox{\fmi X}\,}}   \def\Upsilo{{\mit\Upsilon}}
\def\Ha{{\,\hbox{\fmi H}\,}} \def\mPi{{\mit\Pi}}  \def\tu{{\tau}}
\def\Beta{{\cal B}} \def\bbeta{{\beta}}
\def\sima{{\sigma}}     \def\Ja{{\,\hbox{\fmi J}\,}} 
\def\Ma{{\,\hbox{\fmi M}\,}} \def\Sa{{\,\hbox{\fmi S}\,}}
\def\calJ{{{\cal J}}} \def\calM{{{\cal M}}}
\def\psA{{{\cal A}}} \def\psF{{{\cal F}}}

\def\ag{{\perp\!}}    \def\og{{\eta}}
\def\onab{{\ov\nabla\!}}
\def\Ke{{\,\hbox{\fmi K}\,}} \def\Re{{\,\hbox{\fmi R}\,}}
\def\Fe{{\,\hbox{\fmi F}\,}}  \def\Ce{{\,\hbox{\fmi C}\,}}
\def\Ome{{\mit\Omega}} \def\Xe{{\mit\Xi}} 
\def\ae{{\,\hbox{\fmi a}\,}} 
\def\oDf{{\ov{\cal D}}}  \def\stah{{\star}} 
\def\calE{{\cal E}} \def\oS{\ov{\cal S}}
\def\calC{{\cal C}}    \def\hg{{\gamma}}

\def\ii{{ { i} }}    \def\ji{{ { j}}} 
\def\nauti{{_0}} \def\wuni{{_1}} 
\def\prim{{\prime}} 
\def\dit{{\dot{\,}\!\!}} \def\ddit{{\ddot{\,}\!\!}}
\def\sigme{{\sigma}} \def\Kie{{\,\hbox{\fmi K}\,}}
\def\rhoe{{\rho}} \def\pome{{\varpi}}
\def\lambde{{\lambda}} \def\iote{{\iota}} 
\def\A{{_A}} \def\B{{_B}}
\def\X{{_X}} \def\Y{{_Y}}

\def\phie{{\varphi}} \def\psie{{\psi}}

\def\olag{{\ov{\cal L}}} 
\def\Lambde{{\Lambda}} \def\calK{{\cal K}}
\def\of{{\ov{\,\hbox{\fmi f}}}} 
 \def\tf{{\,\hbox{\fmi f}}} \def\cf{{\check{\,\hbox{\fmi f}}}}
\def\oj{{\ov{\,\hbox{\fmi j}\,}}}  
\def\oW{{\ov{\,\hbox{\fmi j}_{\!_{\{\rm r\}}}\!}}}
\def\bet{{\beta}}  \def\elle{{\ell}} \def\chie{{\chi}}
\def\ch{{\chi}} 
\def\Te{{\,\hbox{\fmi T}\,}} \def\Ue{{\,\hbox{\fmi U}\,}} 
\def\Pe{{\,\hbox{\fmi P}\,}} \def\oT{{\ov{\,\hbox{\fmi T}\,}}} 
\def\cE{{\hbox{\fmi c}_{_{\rm E}}}} \def\cL{{\hbox{\fmi c}_{_{\rm L}}}}
\def\lamb{{\lambda}} \def\varthe{{\vartheta}}
\def\mue{{\mu}} \def\nue{{\nu}}
\def\ue{{\,\hbox{\fmi u}\,}} \def\ve{{\,\hbox{\fmi v}\,}}
\def\calT{{\cal T}} \def \otT{{\widetilde{\,\hbox{\fmi T}\,}}} 
\def\caltT{{\widetilde{\cal T}}} \def\tT{{\ov{\,\hbox{\fmi t}\,}}}
\def\caloT{{\ov{\cal T}}}   \def\pe{{\,\hbox{\fmi p}\,}}  
\def\ce{{\,\hbox{\fmi c}\,}}  \def\ze{{\,\hbox{\fmi z}\,}}
\def\kag{{\mit\Sigma}} \def\crec{{\epsilon}}

\def\hbah{{\hbar}}  
\def\okappa{{\ov  h}}  \def\NG{{\hbox{\fmi G}}} 
\def\Uzero{{{\,\hbox{\fmi U}\,}_{_0}}}  \def\Ye{{\,\hbox{\srm Y}\,}}
\def\Qe{{\,\hbox{\fmi Q}\,}} \def\kappazero{{\kappa_{_0}}} 
\def\ee{{\hbox{\fmi e}}}  \def\qe{{\hbox{\fmi q}}}  
\def\she{{\sharp}}          \def\mast{{{\,\hbox{\fmi m}\,}_\ast}} 
\def\rast{{{\,\hbox{\fmi r}\,}_\ast}}    \def\mag{{\,\hbox{\fmi m}\,}}  
\def\bag{{{\rm b}}}   \def\cag{{{\rm c}}} 
\def\Phie{{\Phi}}    \def\Ac{{\cal I}}
\def\AcI{\spose {\raise 3.0pt\hbox{$\ \acute{\ }$} } {\Ac}}
 \def\Ze{{\,\hbox{\fmi Z}\,}} \def\Ne{{\,\hbox{\fmi N}\,}}

\Title{{Essentials of Classical \\ Brane Dynamics}}

\author{{Brandon Carter}\\
{ D.A.R.C.,
 Observatoire de Paris, 92 Meudon, France }}

\date{{\it Contribution to proc. Peyresq 5 meeting (June, 2000): \\
``Quantum spacetime, brane cosmology,\\
and stochastic effective theories''}}

\maketitle

{\bf Abstract: } This article provides a self contained overview of the
geometry and dynamics of relativistic brane models, of the category that
includes point particle, string, and membrane representations for phenomena
that can be considered as being confined to a worldsheet of the corresponding
dimension (respectively one, two, and three) in a thin limit approximation in
an ordinary 4 dimensional spacetime background. This category also includes
``brane world'' models that treat the observed universe as a 3-brane in 5 or
higher dimensional background. The first sections are concerned with purely
kinematic aspects: it is shown how, to second differential order, the geometry
(and in particular the inner and outer curvature) of a brane worldsheet of
arbitrary dimension is describable in terms of the first, second, and third
fundamental tensor. The later sections show how  -- to lowest order in the
thin limit -- the evolution of such a brane worldsheet will always be governed
by a simple tensorial equation of motion whose left hand side is the
contraction of the relevant surface stress tensor $\oT{^{\mu\nu}}$ with the
(geometrically defined) second fundamental tensor $\Ke_{\mu\nu}{^\rho}$, while
the right hand side will simply vanish in the case of free motion and will
otherwise be just the orthogonal projection of any external force density that
may happen to act on the brane.  

\section{ }

\vfill\eject

\subsection{Introduction}
\label{1-0}

This article is an updated version of the first part of a course originally
presented at a school on ``Formation and Interactions of Topological Defects''
\cite{Carter95}. In preparation for the more specific study of strings in the
later sections, this first part was intended as an introduction to the
systematic study, in a classical relativistic framework, of ``branes", meaning
physical models in which the relevant fields are confined to supporting
worldsheets of lower dimension than the background spacetime. The original
version was motivated mainly by applications in which the background spacetime
dimension was only 4, but the approach described here is particularly
effective for the higher dimensional backgrounds that have very recently
become the subject of intensive investigation by cosmological theorists. 

While not entirely new \cite{Dirac62}, \cite{HoweTucker77}, the development
of classical brane dynamics had lingered at a rather immature stage  (compared
with the corresponding quantum theory \cite{Achucarroetal87} which had been
stimulated by the  rise of ``superstring theory"), the main motivation for
relatively recent work \cite{Carter90} on classical relativistic brane theory
having been its application to vacuum defects produced by the Kibble mechanism
\cite{Kibble76}, particularly when of composite type as in the case of cosmic
strings attached to external domain walls \cite{VilenkinEverett82} and of
cosmic strings carrying internal currents of the kind whose likely existence
was first proposed by Witten \cite{Witten85} and whose potential cosmological
importance, particularly due to the prolific formation of vortons
\cite{DavisShellard89a}, has only recently begun to be generally
recognised \cite{Brandenbergetal96}. However interest in the subject has
suddenly received a substantial boost from an essentially different quarter.

Following the recent incorporation of 10 dimensional ``superstring theory''
into 11 dimensional ``M theory'', the situation has however been radically
changed in the last couple of years by an upsurge 
\cite{ChamGib99,ChamEtal99,BinEtal99} of interest in what has come
to be known as ``brane world'' theory, according to which our observed
4-dimensional universe is to be considered as some kind of brane within a
higher dimensional background that is known in this context as the ``bulk''.
Although they are adequate for cases with codimension 1 (which in the ``brane
world'' context means the most commonly considered case
\cite{ShirEtal00,STW00,GregEtal00,Ms00,MLW00,BatMen00} for 
which the ``bulk'' dimension is only 5) traditional methods of analysis have
been less satisfactory for cases with codimension 2 or more. The advantage,
for such cases, of the more efficient formalism presented here has already
been decisively demonstrated within the framework of an ordinary 4-dimensional
spacetime background, notably in the context of  divergent self interactions
of cosmic strings~\cite{CarterBattye98}) for which previous methods had
provided what turned out to have been misleading results. The superiority of
the present approach should be even more overwhelming for the treatment of
``brane world'' scenarios involving a ``bulk'' having 6 dimensions
\cite{GherghettaShaposhnikov00} or more.

Before the presentation of the generic dynamic laws governing the evolution of
a brane worldsheet (including allowance for the possibility that it may form
the boundary of a higher dimensional brane worldsheet) the first sections of
this article provide a recapitulation of the essential differential geometric
machinery \cite{Carter92a}, \cite{Carter92b} needed for the analysis of a
timelike worldsheet of dimension d say in a background space time manifold of
dimension n. At this stage no restriction will be imposed on the curvature of
the metric -- which will as usual be represented with respect to local
background coordinates $x^\mu$ ($\mu$= 0, ..., n--1) by its components
$\bg_{\mu\nu}$ -- though it will be postulated to be flat, or at least
stationary or conformally flat, in many of the applications to be discussed
later.

\subsection{The first fundamental tensor}
\label{1-1}

The development of geometrical intuition and of computationally efficient
methods for use in string and membrane theory has been hampered by a tradition
of publishing results in untidy, highly gauge dependent, notation (one of the
causes being the undue influence still exercised by Eisenhart's obsolete
treatise ``Riemannian Geometry"  \cite{Eisenhart26}). For the intermediate steps in
particular calculations it is of course frequently useful and often
indispensible to introduce specifically adapted auxiliary structures, such as
curvilinear worldsheet coordinates $\sigme^\ii$ ($\ii$= 0, ..., d--1) and the
associated bitensorial derivatives
\be x^\mu_{\ ,\ii}={\partial x^\mu\over\partial\sigme^\ii}\, ,\eqn{1.1}\fe
or specially adapted orthonormal frame vectors, consisting of an internal 
subset of vectors $\iote_{\!\A}{^\mu}$ ({ $\A$}= 0,  ... , d--1) 
tangential to the worldsheet and an external subset of vectors 
$\lambde_{\X}{^\mu}$ ({ $\X$} = 1, ... , n--d) orthogonal to the 
worldsheet, as characterised by
\be
\iote_{\!\A}{^\mu}\iote_{\!\B\mu}=\eta_{\A\B} \, ,\hskip 1 cm 
\iota_{\!\A}{^\mu}\lambde_{\X\mu}=0\ ,\hskip 1 cm 
\lambde_{\X}{^\mu}\lambde_{\Y\mu}=\delta_{\X\Y}\, ,\eqn{1.2}\fe
where $\eta_{\A\B}$ is a fixed d-dimensional Minkowski metric and the
Kronecker matrix $\delta_{\X\Y}$ is a fixed (n--d)-dimensional Cartesion
metric. Even in the most recent literature there are still (under
Eisenhart's uninspiring influence) many examples of insufficient effort to 
sort out the messy clutter of indices of different kinds (Greek or Latin,
early or late, small or capital) that arise in this way by grouping the
various contributions into simple tensorially covariant combinations. Another
inconvenient feature of many publications is that results have been left 
in a form that depends on some particular gauge choice (such as the 
conformal gauge for internal string coordinates) which obscures the 
relationship with other results concerning the same system but in a 
different gauge.

The strategy adopted here \cite{Stachel80} aims at minimising such problems
(they can never be entirely eliminated) by working as far as possible with a
single kind of tensor index, which must of course be the one that is most
fundamental, namely that of the background coordinates, $x^\mu$. Thus, to
avoid dependence on the internal frame index {$\A$} (which is
lowered and raised by contraction with the fixed d-dimensional Minkowski
metric $\eta_{\A\B}$ and its inverse $\eta^{\A\B}$) and on the external
frame index {$\X$} (which is lowered and raised by contraction with
the fixed (n-d)-dimensional Cartesian metric $\delta_{\X\Y}$ and its inverse
$\delta^{\X\Y}$), the separate internal frame vectors $\iote_{\!\A}{^\mu}$
and external frame vectors $\lambde_{\X}{^\mu}$ will as far as possible be
eliminated in favour of  frame gauge independent combinations
such as the unit tangent d - vector (i.e. antisymmetric contravariant 
d index tensor) with spacetime components given, for a p brane
with p=d -- 1, by
\be \calE^{\mu... \sigma}=({\rm p}+1)!\ \iote_{\!0}{^{[\mu}}
... \iote_{\!{\rm p}}{^{\sigma]}} \, ,\eqn{1.0}\fe
which is useful for many purposes but has the inconvenient feature of being
not strictly tensorial but only pseudo tensorial (since its sign is dependent
on an orientation convention that would be reversed if the ordering of the
frame vectors were subject to an odd permutation) as well as having the 
property (which is particularly awkward for higher dimensional applications) 
that the number of component indices involved is dimension dependent. These 
inconvenient features can however be avoided in many contexts  by
following an approach based on what we refer to as the (first)
fundamental tensor of the worldsheet, which is definable as the (rank d)
operator of tangential projection onto the worldsheet. This
fundamental tensor, which we shall denote here by $\og^\mu_{\ \nu}$,
is given, along with the complementary  (rank n--d) operator 
$\ag^{\!\mu}_{\,\nu}$ of projection orthogonal to the world sheet, by 
\be
\og^\mu_{\ \nu}=\iote_{\!\A}{^\mu}\iote^{\A}{_\nu} \, ,\hskip 1 cm
\ag^{\!\mu}_{\,\nu}=\lambde_{\X}{^\mu}\lambde^{\!\X}{_\nu}\, .\eqn{1.3}\fe

The same principle (of minimisation of the use of auxiliary gauge dependent
reference systems) applies to the avoidance of unnecessary involvement of
the internal coordinate indices which are lowered and raised by contraction
with the induced metric on the worldsheet as given by
\be \hg_{\ii\ji}=\bg_{\mu\nu} x^\mu_{\ ,\ii} x^\nu_{\ ,\ji}\, , \eqn{1.4}\fe
and with its contravariant inverse $\hg^{\ii\ji}$.
After being cast (by index raising if necessary) into its contravariant form,
any internal coordinate tensor can be directly projected onto a corresponding
background tensor in the manner exemplified by the intrinsic metric
itself, which gives 
\be \og^{\mu\nu}= \hg^{\ii\ji} x^\mu_{\ ,\ii} x^\nu_{\ ,\ji} \, ,\eqn{1.5}\fe
thus providing an alternative (more direct) prescription for the fundamental
tensor that was previously introduced via the use of the internal frame in 
(\ref{1.3}). This approach also provides a direct prescription for the orthogonal
projector that was introduced via the use of an external frame in
(\ref{1.3}) but that is also obtainable immediately from (\ref{1.5}) as
\be \ag^{\!\mu}_{\,\nu}=\bg^\mu_{\ \nu}-\og^\mu_{\ \nu}\, .\eqn{1.6}\fe

As well as having the separate operator properties 
\be \og^\mu_{\ \rho}\,\og^\rho_{\ \nu}=\og^\mu_{\ \nu} \ , \hskip 1.2 cm
\ag^{\!\mu}_{\ \rho}\ag^{\!\rho}_{\ \nu}=\ag^{\!\mu}_{\ \nu}
\label{1.6a}\fe 
the tensors defined by (\ref{1.5}) and (\ref{1.6}) will evidently be related
by the conditions
\be \og^\mu_{\ \rho}\ag^{\!\rho}_{\ \nu}\,=\,0\,=\ag^{\!\mu}_{\ \rho}
\og^\rho_{\ \nu} \, . \label{1.6b}\fe

\subsection{The inner and outer curvature tensors}
\label{1-2}

In so far as we are concerned with tensor fields such as the frame vectors
whose support is confined to the d-dimensional world sheet, the effect of
Riemannian covariant differentation $\nabl_\mu$ along an arbitrary
directions on the background spacetime  will not be well defined, only the
corresponding tangentially projected differentiation operation
\be \onab_\mu\eqdef\og\,^\nu_{\ \mu}\nabl_\nu\, , \eqn{1.7}\fe
being meaningful for them, as for instance in the case of a scalar field
$\phie$ for which the tangentially projected gradient is given in
terms of internal coordinate differentiation simply by
$\onab{^\mu}\phie=\hg^{\ii\ji} x^\mu{_{\!,\ii}}\,\phie_{,ji}\,$.

An irreducible basis for the various possible covariant derivatives of 
the frame vectors consists of the { internal rotation} pseudo-tensor
$\rhoe_{\mu\ \rho}^{\,\ \nu}$ and the { external rotation} (or ``twist")
pseudo-tensor $\pome_{\mu\ \rho}^{\,\ \nu}$ as given by
\be \rhoe_{\mu\ \rho}^{\,\ \nu}=\og^\nu\!{_\sigma}\,\iote^{\A}{_\rho}
\onab_\mu\, \iote_{\!\A}{^\sigma}=-\rhoe_{\mu\rho}{^\nu}\, ,\hskip 1 cm
\pome_{\mu\ \rho}^{\,\ \nu}=\ag^{\!\nu}_{\,\sigma}\,\lambde^{\!\X}{_\rho}
\onab_\mu \lambde_{\X}{^\sigma}=-\pome_{\mu\rho}{^\nu}\, , \eqn{1.8}\fe
together with their { mixed} analogue $\Ke_{\mu\nu}{^\rho}$ which is 
obtainable in a pair of equivalent alternative forms given by
\be \Ke_{\mu\nu}{^\rho}=\ag^{\!\rho}_{\,\sigma}\,\iote^{\A}{_\nu}
\onab_\mu\, \iote_{\!\A}{^\sigma}=-\og^\sigma\!{_\nu}\,\lambde_{\X}{^\rho}
\onab_\mu \lambde^{\!\X}{_\sigma}\, .\eqn{1.9}\fe

The reason for qualifying the fields (\ref{1.8}) as ``pseudo" tensors is that
although they are tensorial in the ordinary sense with respect to changes
of the background coordinates $x^\mu$ they are not geometrically well 
defined just by the geometry of the world sheet but are gauge dependent
in the sense of being functions of the choice of the internal and external
frames $\iote_{\!\A}{^\mu}$ and $\lambde_{\X}{^\mu}$. 
The gauge dependence of $\rhoe_{\mu\ \rho}^{\,\ \nu}$ and $\pome_{\mu\
\rho}^{\,\ \nu}$ means that both of them can  be set to zero at any chosen
point on the worldsheet by choice of the relevant frames in its vicinity.
However the condition for it to be possible to set these pseudo-tensors to
zero throughout an open neigbourhood is the vanishing of the curvatures of the
corresponding frame bundles as characterised with respect to the respective
invariance subgroups SO(1,d--1) and SO(n--d) into which the full Lorentz
invariance group SO(1,n--1) is broken by the specification of the
d-dimensional world sheet orientation. The { inner curvature} that needs to
vanish for it to be possible for $\rhoe_{\mu\ \rho}^{\,\ \nu}$ to be set to
zero in an open neighbourhood is of Riemannian type, and is obtainable (by a
calculation of the type originally developed by Cartan that was made familiar
to physicists by Yang Mills theory) as \cite{Carter92a}
\be \Re{_{\kappa\lambda}}{^\mu}{_\nu}=2\og^\mu{_{\!\sigma}}\og^\tau{_{\!\mu}}
\og^\pi{_{[\lambda}}\onab_{\kappa]}\rhoe_{\pi\ \tau}^{\,\ \sigma}+2
\rhoe_{[\kappa}{^{\mu\pi}}\rhoe_{\lambda]\pi\nu}\, , \eqn{1.11}\fe
while the { outer curvature} that needs to vanish for it to be possible
for the ``twist" tensor $\pome_{\mu\ \rho}^{\,\ \nu}$ to be set to zero in 
an open neighbourhood is of a less familiar type that is given \cite{Carter92a} by
\be \Ome{_{\kappa\lambda}}{^\mu}{_\nu}=2\ag^{\!\mu}_{\,\sigma}\ag^{\!\tau}
_{\,\mu}\,\og^\pi{_{[\lambda}}\onab_{\kappa]}\pome_{\pi\ \tau}^{\,\ \sigma}
+2\pome_{[\kappa}{^{\mu\pi}}\pome_{\lambda]\pi\nu}\, .\eqn{1.12}\fe
The frame gauge invariance of the expressions (\ref{1.11}) and (\ref{1.12})
-- which means that  $\Re{_{\kappa\lambda}}{^\mu}{_\nu}$ and 
$\Ome{_{\kappa\lambda}}{^\mu}{_\nu}$ are unambiguously well defined as
tensors in the strictest sense of the word -- is not immediately obvious from
the foregoing formulae, but it is made manifest in the the alternative
expressions given in Subsection \ref{1-5}.

\subsection{The second fundamental tensor}
\label{1-3}

Another, even more fundamentally important, gauge invariance property that is
not immediately obvious from the traditional approach -- as recapitulated in
the preceeding subsection is -- that of the entity  $ \Ke_{\mu\nu}{^\rho}$
defined by the mixed analogue (\ref{1.9}) of (\ref{1.8}), which  (unlike
$\rhoe_{\mu\ \rho}^{\,\ \nu}$ and $\pome_{\mu\ \rho}^{\,\ \nu}$, but like
$\Re{_{\kappa\lambda}}{^\mu}{_\nu}$ and $\Ome{_{\kappa\lambda}}{^\mu}{_\nu}$)
is in fact a geometrically well defined tensor in the strict sense. To see
that the formula (\ref{1.9}) does indeed give a result that is frame gauge
independent, it suffices to verify that it agrees with the alternative --
manifestly gauge independent definition  \cite{Carter90} \be
\Ke_{\mu\nu}{^\rho} \eqdef \og\,^\sigma_{\ \nu}\onab_\mu \og\,^\rho_{\ \sigma}
\, .\eqn{1.10}\fe whereby the entity that we refer to as the { second
fundamental tensor} is constructed directly from the the first fundamental
tensor $\og^{\mu\nu}$ as given by (\ref{1.5}).

Since this second fundamental tensor, $\Ke_{\mu\nu}{^\rho}$ will play a very
important role throughout the work that follows, it is worthwhile to linger
over its essential properties. To start with it is to be noticed that a
formula of the form (\ref{1.10}) could of course be meaningfully meaningful
applied not only to the fundamental projection tensor of a d-surface, but also
to any (smooth) field of rank-d projection operators $\og\,^\mu_{\ \nu}$ as
specified by a field of arbitrarily orientated d-surface elements. What
distinguishes the integrable case, i.e. that in which the elements mesh
together to form a well defined d-surface through the point under
consideration, is the condition that the tensor defined by (\ref{1.10}) should
also satisfy the { Weingarten identity} 
\be \Ke_{[\mu\nu]}{^\rho} =0 \eqn{1.13}\fe 
(where the square brackets denote antisymmetrisation), this
symmetry property of the second fundamental tensor being
derivable \cite{Carter90}, \cite{Carter92a} as a version of the well known Frobenius
theorem. In addition to this non-trivial symmetry property, the second
fundamental tensor is also obviously tangential on the first two indices and
almost as obviously orthogonal on the last, i.e. 
\be\ag^{\!\sigma}_{\,\mu}\Ke_{\sigma\nu}{^\rho}=\Ke_{\mu\nu}{^\sigma}
\og_\sigma{^\rho}=0   \, . \eqn{1.14}\fe
The second fundamental tensor $\Ke_{\mu\nu}{^\rho}$ has the property of fully
determining the tangential derivatives of the first fundamental tensor
$\og\,^\mu_{\ \nu}$ by the formula
\be \onab_\mu\og{_{\nu\rho}}=2\Ke_{\mu(\nu\rho)} \eqn{1.15}\fe
(using round brackets to denote symmetrisation) and it can be seen to be
characterisable by the condition that the orthogonal projection of the
acceleration of any tangential unit vector field $\ue^\mu$ will be given by
\be \ue^\mu \ue^\nu \Ke_{\mu\nu}{^\rho}=\ag^{\!\rho}_{\,\mu}\dot \ue^\mu\ ,
\hskip 1 cm \dot \ue^\mu={\ue^\nu\nabl{_\nu} \ue^\mu} \, . \eqn{1.16}\fe

In cases for which we need to use the d index surface element 
pseudo tensor $\calE^{\mu... \sigma}$ given for the d dimensional  
worldsheet of the p brane by (\ref{1.0}), it will be useful to have 
the relevant surface derivative formula which takes the form
\be \onab_\lambda \calE^{\mu... \sigma}= (-1)^{\rm p}\, ({\rm p}+1)\
\calE^{\nu[\mu ...}\Ke_{\lambda\nu}{^{\sigma]}} \, ,\eqn{0.0}\fe
in which it is to be recalled that p=d--1 .
(This expression corrects what is, as far as I am aware, the only
wrongly printed formula in the more complete analysis \cite{Carter92a} on 
which this presentation is based: the factor $(-1)^{\rm p}$ was
indvertently omitted in the relevant formula (B9), which is thus
valid as printed only for a worldsheet of odd dimension d=p+1.)

\subsection{Extrinsic curvature vector and conformation tensor} 
\label{1-4}

It is very practical for a great many purposes to introduce the {
extrinsic curvature vector} $K^\mu$, defined as the trace of the second
fundamental tensor, which is automatically orthogonal to the worldsheet, 
\be \Ke^\mu\eqdef \Ke^\nu_{\ \nu}{^\mu}
\ , \hskip 1 cm  \og^\mu_{\ \nu}\Ke {^\nu}=0 \, .\eqn{1.17}\fe
It is useful for many specific purposes to work this out in terms of the
intrinsic metric $\hg_{\ii\ji}$ and its determinant $\vert\hg\vert $. 
It suffices to use the simple expression 
$\onab^{\,\mu}\phie=\hg^{\ii\ji} x^\mu{_{,\ii}}\phie_{,\ji}$ 
for the tangentially projected gradient of a scalar field $\phie$ on the
worldsheet, but for a tensorial field (unless one is using Minkowski
coordinates in a flat spacetime) there will also be contributions involving
the background Riemann Christoffel connection 
\be\Gamma_{\mu\ \rho}^{\,\ \nu}=\bg^{\nu\sigma}\big(\bg_{\sigma(\mu,\rho)}
-{_1\over^2}\bg_{\mu\rho,\sigma}\big)\, .\eqn{1.18}\fe
The curvature vector is thus obtained in explicit detail as
\be \Ke^\nu=\onab_\mu\og^{\mu\nu}={1\over\sqrt{\Vert \hg\Vert}}\Big(
\sqrt{\Vert\hg\Vert}\hg^{\ii\ji}x^\nu_{\, ,\ii}\Big){_{,\ji}}+
\hg^{\ii\ji}x^\mu_{\, ,\ii}x^\rho_{\, ,\ji}\Gamma_{\mu\ \rho}^{\,\ \nu}\, .
\eqn{1.19}\fe
This last expression is  technically useful for certain specific
computational purposes, but it must be remarked that much of the 
literature on cosmic string dynamics has been made unnecessarily heavy 
to read by a tradition of working all the time with long strings of non
tensorial terms such as those on the right of (\ref{1.19}) rather than taking
advantage of such more succinct tensorial expressions as the preceeding 
formula $\onab_\mu\og^{\mu\nu}$. As an alternative to the universally 
applicable tensorial approach advocated here, there is of course another more 
commonly used method of achieving succinctness in particular circumstances,
which is to sacrifice gauge covariance by using specialised kinds of 
coordinate system. In particular for the case of a string, i.e. for a 
2-dimensional worldsheet, it is standard practise to use conformal 
coordinates $\sigme^{\nauti}$ and $\sigme^{\wuni}$ so that the corresponding 
tangent vectors $\dit x^\mu=x^\mu_{\, ,\nauti}$ and $x^{\prim\mu}= 
x^\mu_{\, ,\wuni}$ satisfy the restrictions $\dit x^\mu x^\prim_{\, \mu}=0$, 
$\dit x^\mu\dit x_\mu+x^{\prim\mu}x^\prim_{\,\mu}=0$, which implies 
$\sqrt{\Vert \hg\Vert}=x^{\prim\mu}x^\prim_{\,\mu}=-\dit x^\mu\dit x_\mu$ 
so that (\ref{1.19}) simply gives $\sqrt{\Vert\hg \Vert}\,\Ke^\nu=$ 
$x^{\prim\prim\nu}-\ddit x^\nu + (x^{\prim\mu}x^{\prim\rho}
-\dit x^\mu\dit x^\rho)\Gamma_{\mu\ \rho}^{\,\ \nu}$.

The physical specification of the extrinsic curvature vector (\ref{1.17}) for
a timelike d-surface in a dynamic theory provides what can be taken as the
equations of extrinsic motion of the d-surface \cite{Carter90,Carter92b}, the
simplest possibility being the ``harmonic" condition $\Ke^\mu=0$ that is
obtained (as will be shown in the following sections) from a surface measure
variational principle such as that of the Dirac membrane model \cite{Dirac62},
or of the Goto-Nambu string model \cite{Kibble76} whose dynamic equations in a
flat background are therefore expressible with respect to a standard conformal
gauge in the familiar form $x^{\prim\prim\mu}-\ddit x^\mu=0$. 

There is a certain analogy between the Einstein vacuum equations, which
impose the vanishing of the trace $\calR_{\mu\nu}$ of the background
spacetime curvature $\calR_{\lambda\mu}{^\rho}{_\nu}$, and the
Dirac-Gotu-Nambu equations, which impose the vanishing of the trace
$\Ke^\nu$ of the second fundamental tensor $\Ke_{\lambda\mu}{^\nu}$.
Just as it is useful to separate out the Weyl tensor
\cite{Schouten54}, i.e. the trace free part of the Ricci background
curvature which is the only part that remains when the Einstein vacuum
equations are satisfied, so also analogously, it is useful to separate
out the the trace free part of the second fundamental tensor, namely
the extrinsic conformation tensor \cite{Carter92a}, which is the only
part that remains when equations of motion of the Dirac - Goto - Nambu type
are satisfied. Explicitly, the trace free { extrinsic conformation} tensor
$\Ce_{\mu\nu}{^\rho}$ of a d-dimensional imbedding is defined \cite{Carter92a}
in terms of the corresponding first and second fundamental tensors
$\eta_{\mu\nu}$ and $\Ke_{\mu\nu}{^\rho}$ as \be \Ce_{\mu\nu}{^\rho}\eqdef
\Ke_{\mu\nu}{^\rho}-{1\over{\rm d}}\og{_{\mu\nu}} \Ke^\rho \ , \hskip 1 cm
\Ce^\nu_{\ \nu}{^\mu}=0 \ .\eqn{1.20}\fe Like the Weyl tensor
$\calW_{\lambda\mu}{^\rho}{_\nu}$ of the background metric (whose  definition
is given implicitly by (\ref{1.25}) below) this conformation tensor has the
noteworthy property of being invariant with respect to conformal modifications
of the background metric: \be \bg_{\mu\nu}\mapsto {\rm
e}^{2\alpha}\bg_{\mu\nu}\, ,\hskip 0.6 cm \Rightarrow\hskip 0.6 cm
\Ke_{\mu\nu}{^\rho}\mapsto \Ke_{\mu\nu}{^\rho}
+\eta_{\mu\nu}\ag^{\!\rho\sigma}\nabl_\sigma\alpha\, ,\hskip 0.6 cm
\Ce_{\mu\nu}{^\rho}\mapsto \Ce_{\mu\nu}{^\rho}\, .\eqn{1.21}\fe This formula
is useful \cite{Carteretal94} for calculations of the kind undertaken by
Vilenkin \cite{Vilenkin91} in a standard Robertson-Walker type cosmological
background, which can be obtained from a flat auxiliary spacetime metric by a
conformal transformation for which ${\rm e}^\alpha$ is a time dependent Hubble
expansion factor. 

\subsection{The Codazzi, Gauss, and Schouten identities}
\label{1-5}

As the higher order analogue of (\ref{1.10}) we can go on to introduce the 
{ third} fundamental tensor$ \cite{Carter90}$ as
\be \Xe_{\lambda\mu\nu}{^\rho} \eqdef\og\,^\sigma_{\ \mu}\og\,^\tau_{\ \nu}
\ag^{\!\rho}_{\,\alpha}\onab_\lambda \Ke_{\sigma\tau}{^\alpha} \, , 
\eqn{1.22}\fe
which  by construction is obviously symmetric between the second and third
indices and tangential on all  the first three indices.  In a spacetime 
background that is flat (or of constant curvature as is the case for the 
DeSitter universe model) this third fundamental tensor is fully symmetric 
over all the first three indices by what is interpretable as the {
generalised Codazzi identity} which is expressible \cite{Carter92a} in a 
background with arbitrary Riemann curvature $\calR_{\lambda\mu}{^\rho}{_\sigma}$ 
as 
\be \Xe_{\lambda\mu\nu}{^\rho}= \Xe_{(\lambda\mu\nu)}{^\rho} +{_2\over^3}
\og\,^\sigma_{\ \lambda}\og\,^\tau_{\ {(\mu}}  \og\,^\alpha_{\ {\nu)}}
\calR_{\sigma\tau}{^\beta}{_\alpha}\ag^{\!\rho}_{\,\beta}
\  \eqn{1.23}\fe
It is to be noted that a script symbol $\calR$ is used here in order to
distinguish the (n - dimensional) background Riemann curvature tensor from the
intrinsic curvature tensor (\ref{1.11}) of the (d - dimensional) worldship to
which the ordinary symbol $\Re$ has already allocated.

For many of the applications that will follow it will be sufficient just to
treat the background spacetime as flat, i.e. to take
$\calR_{\sigma\tau}{^\beta}{_\alpha}=0$. At this stage however, we shall allow
for an unrestricted background curvature. For n$>2$ this will be decomposible
in terms of its trace free  Weyl part $\calW_{\mu\nu}{^\rho}{_\sigma}$ (which as
remarked above is conformally invariant) and the corresponding background
Ricci tensor and its scalar trace,
\be \calR_{\mu\nu}= \calR_{\rho\mu}{^\rho}{_\nu}  \, , \hskip 1 cm
 \calR=\calR^\nu_{\ \nu}\, , \eqn{1.24}\fe
in the form  \cite{Schouten54}
\be\calR_{\mu\nu}{^{\rho\sigma}}=\calW_{\mu\nu}{^{\rho\sigma}} +{_4\over^{ n-2} }
\bg^{[\rho}_{\ [\mu}\calR^{\sigma]}_{\ \nu]}-{_2\over ^{(n-1)(n-2)} } \calR
\bg^{[\rho}_{\ [\mu}\bg^{\sigma]}_{\ \nu]} \, ,\eqn{1.25}\fe
(in which the Weyl contribution can be non zero only for n$\geq$ 4). In terms
of the tangential projection of this background curvature, one can evaluate
the corresponding { internal} curvature tensor (\ref{1.11}) in the form
\be  \Re{_{\mu\nu}}{^\rho}{_\sigma}= 2\Ke^\rho{_{[\mu}}{^\tau}
\Ke_{\nu]\sigma\tau}+ \og\,^\kappa_{\ \mu} \og\,^\lambda_{\ \nu}
\calR_{\kappa\lambda}{^\alpha}{_\tau} \og\,^\rho_{\ \alpha}\og\,^\tau_{\ \sigma}
  \, , \eqn{1.26}\fe
which is the translation into the present scheme of what is well known in
other schemes as the { generalised Gauss identity}. The much less well
known analogue for the (identically trace free and conformally invariant) {
outer} curvature (\ref{1.12}) (for which the most historically appropriate
name might be argued to be that of Schouten \cite{Schouten54}) is given 
\cite{Carter92a}in terms of the corresponding projection of the background 
Weyl tensor by the expression 
\be \Ome{_{\mu\nu}}{^\rho}{_\sigma}= 2\Ce_{[\mu}{^{\tau\rho}}
\Ce_{\nu]\tau\sigma}+ \og\,^\kappa_{\ \mu} \og\,^\lambda_{\ \nu}
\calW_{\kappa\lambda}{^\alpha}{_\tau}\ag^{\!\rho}_{\,\alpha}
\ag^{\!\tau}_{\,\sigma} \, . \eqn{1.27}\fe
It follows from this last identity  that in a background that is flat or 
conformally flat (for which it is necessary, and for n$\geq 4$ sufficient, that
the Weyl tensor should vanish) the vanishing of the extrinsic conformation
tensor $\Ce_{\mu\nu}{^\rho}$ will be sufficient (independently of the
behaviour of the extrinsic curvature vector $\Ke^\mu$) for vanishing of the
outer curvature tensor $\Ome{_{\mu\nu}}{^\rho}{_\sigma}$, which is the
condition for it to be possible to construct fields of vectors $\lambde^\mu$
orthogonal to the surface and such as to satisfy the generalised
Fermi-Walker propagation  condition to the effect that $\ag^{\!\rho}_{\,\mu}
\onab_\nu\lambde_\rho$ should vanish. It can also be shown \cite{Carter92a}
(taking special trouble for the case d=3 )  that in a conformally flat
background (of arbitrary dimension n) the vanishing of the conformation
tensor $\Ce_{\mu\nu}{^\rho}$ is always sufficient (though by no means
necessary) for conformal flatness of the induced geometry in the imbedding.

\subsection{ The Internal Ricci and Conformal Curvatures.}
\label{1-5a}

The conclusion of the preceding paragraph is an illustration of the 
critically significant role of the conformation tensor 
$\Ce{_{\mu\nu}{^\rho}}$ of an imbedding when the background is 
conformally flat, which suggests that it will be of interest to make 
a closer examination of its role with respect to the { inner} 
curvature, $\Re_{\kappa\lambda}{^\mu}{_\nu}$ and more particularly of its
tensorially irreducible parts, in this conformally flat case, for which  
the condition that the background Weyl tensor should vanish
is necessary -- and for $n\geq 4$ also sufficient \cite{Schouten54} --
while when the background dimension is $n=3$ this condition, namely
 $\calW_{\kappa\lambda}{^\mu}{_\nu}= 0 $, will
will hold in any case as an identity. This restriction is of course 
compatible with all the most common kinds of application, in which the 
background is taken to be not just conformally flat, but flat in the 
strong sense, which is justifiable at least as a very good approximation 
in a very wide range of circumstances in which the characteristic length 
scales of the imbedding will be small compared with those of the 
background curvature if any. Although it is unnecessary for such cases, 
we shall nevertheless retain allowance for the possibility of a non zero 
background Ricci tensor $\calR_{\mu\nu}$ in the formulae that follows since the
extra complication involved thereby is only very moderate (compared with 
what would result if allowance for a non zero background Weyl tensor were 
also included).

Leaving aside the trivial (always locally conformally flat) case of a 
2-dimensional background, the generalised Gauss relation (\ref{1.26}) 
reduces to the  form
$$ \Re_{\kappa\lambda}{^\mu}{_\nu}={2\over n-2} \big(\og_{[\kappa}{^{\mu}} \og_{\lambda]}{^\rho}
\og_\nu{^\sigma}-\og_{\nu[\kappa}\eta_{\lambda]}{^\rho}\og^{\mu\sigma}
\big) \Big(\calR_{\rho\sigma}-{\calR \over 2(n-1)}\bg_{\rho\sigma}\Big)$$
\be \hskip 1 cm +2\Ke_{[\kappa}{^{\mu\sigma}}\Ke_{\lambda]
\nu\sigma} +
\og_\kappa{^\rho} \og_\lambda{^\sigma}
\calW_{\rho\sigma}{^\tau}{_\upsilon}
 \og^\mu{_\tau} \og^\upsilon{_\nu} \,  , \eqn{(31)}\fe
in which the last term evidently drops out whenever the background Weyl 
tensor vanishes. Proceeding from this formula by contraction, the
internal Ricci tensor is obtained in terms of  the irreducible parts
$\Ke_\rho$ and $\Ce_{\lambda\mu}{^\nu}$ of the second fundamental tensor
$\Ke{_{\mu\nu}}{^\rho}$
in the form
\be  \Re_{\mu\nu}={p-2\over n-2}\og_\mu{^\rho}\eta_\nu{^\sigma}
\calR_{\rho\sigma}+{1\over n-2}\left( \og^{\rho\sigma} \calR_{\rho\sigma}
-{p-1\over n-1}\calR \right)  \og_{\mu\nu} \hskip 1 cm   $$ $$\hskip 1.5 cm
+{p-1\over p^2} \Ke^\sigma \Ke_\sigma   \og_{\mu\nu}
+{p-2\over p} \Ce_{\mu\nu}{^\sigma} \Ke_\sigma
-\Ce_\mu{^{\rho\sigma}}\Ce_{\nu\rho\sigma} 
+\calW_{\mu\nu}\ ,   \eqn{(32)}\fe
where the final background Weyl contribution, if any, is given by
the expressions
\be \calW_{\mu\nu}=\og_\mu{^\sigma} \og_\nu{^\kappa}\,
\calW_{\rho\sigma}{^\tau}{_\kappa}
 \og^\rho{_\tau}=-\og_\mu{^\sigma} \og_\nu{^\kappa}\,
\calW_{\rho\sigma}{^\tau}{_\kappa}  \ag^\rho{_\tau}
 \, ,\eqn{(33)}\fe
of which the last version is obtained as a consequence of 
the tracelessnes of the Weyl tensor.

The corresponding Ricci scalar for the internal geometry (whose surface 
integral in the special case $p=2$ gives the ordinary Gauss Bonnet type 
invariant that was mentionned at the end of section 8) is thus
finally obtained in the form
\be  \Re ={p-1\over n-2}\left(2 \og^{\rho\sigma} \calR_{\rho\sigma}-
{p\over n-1} \calR \right) + {p-1\over p}\Ke^\sigma \Ke_\sigma -
\Ce_{\lambda\mu}{^\nu}\Ce^{\lambda\mu}{_\nu}+\calW\ , \eqn{(34)}\fe
(which corrects a transcription error whereby a factor of two
was omitted in the original version \cite{Carter92a})
where the final Weyl contribution is just the trace
\be \calW=\calW^\nu{_\nu}=\og^{\rho\tau}\calW_{\rho\sigma}{^\tau}{_\nu}
\og^\nu{_\tau}=\ag^{\rho\tau}\calW_{\rho\sigma}{^\tau}{_\nu}
\ag^\nu{_\tau}\, ,\eqn{(35)}\fe
which can be seen to vanish identically unless both the dimension
and the codimension of the worldsheet are greater than one,
i.e. unless both $p\geq 2$ and $n-p\geq 2$.

For cases in which the imbedded surface has dimension $p\leq 3$, as must 
always be the case in an ordinary 4-dimensional space-time background, the
specification of the Ricci contribution provides all that is needed to 
specify the complete inner curvature tensor. However to fully specify
$\Re_{\kappa\lambda}{^\mu}{_\nu}$ in higher dimensional cases for which the
imbedded surface has dimension $p\geq 4$ it will also be necessary to 
take account of the generically non zero conformal curvature term
$\calC_{\kappa\lambda}{^\mu}{_\nu}$ that will contribute to the total as given 
by the internal analogue of (\ref{1.25}), namely
\be\Re_{\mu\nu}{^{\rho\sigma}}=\calC_{\mu\nu}{^{\rho\sigma}} +{_4\over^{ p-2} }
\og^{[\rho}_{\ [\mu}\Re^{\sigma]}_{\ \nu]}-{_2\over ^{(p-1)(p-2)} } \Re
\og^{[\rho}_{\ [\mu}\og^{\sigma]}_{\ \nu]} \, .\eqn{36}\fe

 The rather greater algebraic effort  required to work out this 
inner conformal curvature contribution is rewarded by the qualitatively tidy 
form of the result, which (in contrast with the miscellaneous form of the 
terms assembled in (\ref{(32)}) and (\ref{(34)}) is homogeneously quadratic in the 
conformation tensor alone, the contributions of the trace vector $\Ke^\mu$ and 
of the background Ricci tensor $\calR_{\mu\nu}$  again (as in (\ref{1.27})) being 
found to miraculously cancel out altogether, leaving
$$ \calC_{\kappa\lambda}{^{\mu\nu}}=2\Ce_{[\kappa}{^{\mu\sigma}}
 \Ce_{\lambda]}{^\nu}{_\sigma}  
  -{_4\over ^{p-2}}\Big( \Ce^{\rho[\mu}{_\sigma}
\og^{\nu]}{_{[\kappa}}\Ce_{\lambda]\rho}{^\sigma} +\og_{[\kappa}{^{[\mu}}
\calW_{\lambda]}{^{\nu]}}\Big) $$ \be
  - {_2\over ^{(p-2)(p-1)}}
\og_{[\kappa}{^\mu}\og_{\lambda]}{^\nu}\Big( \Ce_{\rho\sigma}{^\tau}
\Ce^{\rho\sigma}{_\tau} -\calW\Big)
+\og_\kappa{^\rho} \og_\lambda{^\sigma}
\calW_{\rho\sigma}{^\tau}{_\upsilon}\og^\mu{_\tau} \og^{\upsilon\nu} 
\ . \eqn{(36)}\fe
 We can thus draw the memorable conclusion that in a conformally
flat background the vanishing of the
conformation tensor $\Ce^{\mu\nu}{_\rho}$ is a sufficient condition not 
only for (local) outer flatness but also for (local) internal conformal 
flatness, at least for an imbedded surface with dimension $p\geq 4$.
With a little more work \cite{Carter92a} it can be shown that
this conclusion also holds for $p=3$, while it is trivial
for the case of a string worldheetm $p=2$, which is always
(locally) conformally flat.

\subsection{The special case of a string worldsheet in 
4-dimensions} \label{1-8}

The application with which we shall mainly be concerned in the following work
will be the case d=2 of a string. An orthonormal tangent frame will consist in
this case just of a timelike unit vector, $\iote\nauti{^\mu}$, and a spacelike
unit vector, $\iote\wuni{^\mu}$, whose exterior product vector is the frame
independent antisymmetric unit surface element tensor 
\be \calE^{\mu\nu}=2\iote\nauti{^{[\mu}}\iote\wuni{^{\nu]}}
=2\big(\!-\!\vert\hg\vert\big)^{-1/2}\,x^{[\mu}_{\ \, ,\nauti}x^{\nu]}_{\ ,\wuni}
\, ,\eqn{1.28}\fe
whose tangential gradient satisfies
\be \onab_\lambda\calE^{\mu\nu}=-2\Ke_{\lambda\rho}{^{[\mu}}\calE^{\nu]\rho}\, .
\eqn{1.29}\fe
In this case the inner rotation pseudo
tensor (\ref{1.8}) is determined just by a corresponding rotation covector
$\rhoe_\mu$ according to the specification
\be \rhoe_{\lambda\ \nu}^{\ \mu}={_1\over^2}\,\calE^\mu_{\ \nu}\rhoe_\lambda\, ,
\hskip 1 cm \rhoe_\lambda=\rhoe_{\lambda\ \nu}^{\ \mu}\calE^\nu_{\ \mu}  
\, .\eqn{1.30}\fe
This can be used to see from (\ref{1.11}) that the Ricci scalar,
\be \Re = \Re{^\nu_{\ \nu}}\, 
\hskip 1 cm \Re{_{\mu\nu}}=\Re{_{\rho\mu}}{^\rho}{_\nu}
\, ,\eqn{1.31a}\fe
of the 2-dimensional worldsheet will have the well known property
of being a pure surface divergence, albeit of a frame gauge dependent 
quantity:
\be \Re=\onab_\mu\big(\calE^{\mu\nu} \rhoe_\nu\big) \, .\eqn{1.31b}\fe
In the specially important case of a string in ordinary 4-dimensional
spacetime, i.e. when we have not only d=2 but also n=4, the antisymmetric
background measure tensor $\bepsilon^{\lambda\mu\nu\rho}$ can be used to
determine a scalar (or more strictly, since its sign is orientation
dependent, a pseudo scalar) magnitude $\Ome$ for the outer curvature
tensor (\ref{1.12}) (despite the fact that its traces are identically zero)
according to the specification
\be \Ome={_1\over^2}\,\Ome_{\lambda\mu\nu\rho}\,\bepsilon^
{\lambda\mu\nu\rho}\, .\eqn{1.32}\fe
Under these circumstances one can also define a ``twist" covector
$\pome_\mu$, that is the outer analogue of $\rhoe_\mu$, according to the
specification
\be \pome_\nu={_1\over^2}\,\pome_\nu^{\ \mu\lambda}\,
\bepsilon_{\lambda\mu\rho\sigma}\,\calE^{\rho\sigma}\, .\eqn{1.33}\fe
This can be used to deduce from (\ref{1.12}) that the outer curvature (pseudo)
scalar $\Omega$ of a string worldsheet in 4-dimensions has a divergence
property of the same kind as that of its more widely known Ricci analogue
(\ref{1.31b}), the corresponding formula being given by
\be \Ome=\onab_\mu\big(\calE^{\mu\nu} \pome_\nu\big) \ .\eqn{1.34}\fe
It is to be remarked that for a compact spacelike 2-surface the integral of
(\ref{1.29}) gives the well known Gauss Bonnet invariant, but that the
timelike string worldsheets under consideration here will not be characterised
by any such global invariant since they will not be compact (being open in the
time direction even for a loop that is closed in the spacial sense). The outer
analogue of the Gauss Bonnet invariant that arises from (\ref{1.32}) for a
spacelike 2-surface has been discussed by Penrose and Rindler 
\cite{PenroseRindler84} but again there is no corresponding global invariant 
in the necessarily non-compact timelike case of a string worldsheet.

\subsection{Regular and distributional formulations of brane action} 
\label{2-1}

The term p-brane has come into use \cite{Achucarroetal87}, 
\cite{BarsPope88} to describe a dynamic
system localised on a timelike support surface of dimension d=p+1, 
imbedded in a spacetime background of dimension n$>$p. Thus at the
low dimensional extreme one has the example of a zero - brane, meaning what
is commonly referred to as a ``point particle", and of a 1-brane meaning
what is commonly referred to as a ``string". At the high dimensional extreme
one has the ``improper" case of an (n--1)-brane, meaning what is commonly
referred to as a ``medium" (as exemplified by a simple fluid), and of an
(n--2)-brane, meaning what is commonly referred to as a ``membrane" (from which 
the generic term ``brane" is derived). A membrane (as exemplified by a
cosmological domain wall) has the special feature of being supported by a
hypersurface, and so being able to form a boundary between separate
background space time regions; this means that a 2-brane has the status of
being a membrane in ordinary 4-dimensional spacetime (with $n=4$) but not in
a higher dimensional (e.g. Kaluza Klein type) background.

The purpose of the present section is to consider the dynamics not just of
an individual brane but of a { brane complex} or ``rigging model" 
\cite{Carter90}
such as is illustrated by the nautical archetype in which the wind -- a
3-brane -- acts on a boat's sail -- a 2-brane -- that is held in place by
cords -- 1-branes -- which meet at knots, shackles and pulley blocks that
are macroscopically describable as point particles -- i.e. 0-branes. In
order for a a set of branes of diverse dimensions to qualify as
a``geometrically regular'' brane complex or ``rigging system" it is required
not only that the support surface of each (d--1)-brane should be a smoothly
imbedded d-dimensional timelike hypersurface but also that its boundary, if
any, should consist of a disjoint union of support surfaces of an attatched
subset of lower dimensional branes of the complex. (For example in order
qualify as part of a regular brane complex the edge of a boat's sail can not
be allowed to flap freely but must be attatched to a hem cord belonging to
the complex.) For the brane complex to qualify as regular in the strong
dynamic sense that will be postulated in the present work, it is also
required that a member p-brane can exert a direct force only on an an
attached (p--1)-brane on its boundary or on an attached (p+1)-brane on whose
boundary it is itself located, though it may be passively subject to forces
exerted by a higher dimensional background field. For instance the
Peccei-Quin axion model gives rise to field configurations representable as
regular complexes of domain walls attached to strings 
\cite{VilenkinEverett82,Sikivie82,Shellard90}, and a bounded (topological or
other) Higgs vortex defect terminated by a pair of pole defects 
\cite{Nambu77,Manton83,Copelandetal88a,VachAch91,VachBar92,MartinVilenkin96} 
may be represented as a regular brane complex consisting of a finite cosmic
string with a pair of point particles at its ends, in an approximation
neglecting Higgs field radiation. (However allowance for radiation would 
require the use of an extended complex including the Higgs medium whose 
interaction with the string -- and a fortiori with the terminating particles 
-- would violate the regularity condition: the ensuing singularities in the 
back reaction would need to be treated by a renormalisation procedure of a
kind \cite{Shellard90,DabholkarQuashnock90,BattyeShellard95,BattyeShellard96}
whose development so far has been beset with difficulties in preserving exact
local Lorentz invariance, an awkward problem that is beyond the scope of the
present article.)

The present section will be restricted to the case of a brane complex that
is not only regular in the sense of the preceeding paragraph but that is
also { pure} (or ``fine") in the sense that the lengthscales
characterising the internal structure of the (defect or other) localised
phenomenon represented by the brane models are short compared with those
characterising the macroscopic variations under consideration so that
polarisation effects play no role. For instance in the case of  a point
particle, the restriction that it should be describable as a ``pure" zero
brane simply means that it can be represented as a simple monopole without
any dipole or higher multipole effects. In the case of a cosmic string the
use of a ``pure" 1-brane description requires that the underlying vortex
defect be sufficiently thin compared not only compared with its total length
but also compared with the lengthscales characterising its curvature
and the gradients of any currents it may be carrying. The effect of the
simplest kind of curvature corrections beyond this ``pure brane" limit 
has been considered by several authors for strings 
\cite{Po86,MT88,G88,G93}, domain walls 
\cite{Gal91,SM93,Bal94,CGr95}, and more generally
\cite{Let90,HT90,A92,BL92,C94,CGu95}, but in the rest of this article, 
as in the present section, it will be assumed that the ratio of microscopic 
to macroscopic lengthscales is sufficiently small for description in terms 
of ``pure" p-branes to be adequate.

The present section will not be concerned with the specific details of
particular cases but with the generally valid laws that can be derived as
Noether identities from the postulate that the model is governed by dynamical
laws derivable from a variational principle specified in terms of an action
function ${\Ac}$. It is however to be emphasised that the validity at a
macroscopic level of the laws given here is not restricted to cases
represented by macroscopic models of the strictly conservative type directly
governed by a macroscopic variational principle. The laws obtained here will
also be applicable to classical models of dissipative type (e.g. allowing for
resistivity to relative flow by internal currents) as necessary conditions for
the existence of an underlying variational description  of the microscopic
(quantum) degrees of freedom that are allowed for merely as entropy in the
macroscopically averaged classical description.

In the case of a brane complex, the total action $\Ac$ will be given as a sum
of distinct d-surface integrals respectively contributed by the various
(d--1)-branes of the complex, of which each is supposed to have its own
corresponding Lagrangian surface density scalar $\ud \olag$ say. Each supporting
d-surface will be specified by a mapping $\sigme\mapsto x\{\sigme\}$ giving
the local background coordinates $x^\mu$ ($\mu$=0, .... , n--1) as functions
of local internal coordinates $\sigme^\ii\ $ ( i=0, ... , d--1). The
corresponding d-dimensional surface metric tensor $\ud \hg_{\ii\ji}$ that is
induced (in the manner described in Subsection \ref{1-1}) as the pull back
of the n-dimensional background spacetime metric $\bg_{\mu\nu}$, will
determine the natural surface measure, $\ud d\!\oS$, in terms of 
which the total action will be expressible in the form 
\be \Ac= \sum_{\rm d} \int\!\! \ud d\!\oS\, \ud \olag \ ,
 \hskip 1 cm \ud d\!\oS=\sqrt{\Vert^{_{\rm (d)}} \hg \Vert}\,
 d^{\rm d}\!\sigme \, . \eqn{2.1}\fe
As a formal artifice whose use is an unnecessary complication in ordinary
dynamical calculations but that can be useful for  purposes such as the
calculation of radiation, the { confined} (d-surface supported) but
locally { regular} Lagrangian scalar fields $\ud \olag$ can be
replaced by corresponding unconfined, so no longer regular but {
distributional} fields $\ud \hlag$, in order to allow the the basic
multidimensional action (\ref{2.1}) to be represented as a single integral, 
\be \Ac =\int\!\! d\!{\calS}\, \sum_{\rm d} \ud \hlag \ ,
 \hskip 1 cm d\!{\calS}=\sqrt{\Vert \bg \Vert}\, d^{\rm n}x \, .\eqn{2.2}\fe
over the n-dimensional { background} spacetime. In order to do this, it is
evident that for each (d--1)-brane of the complex the required distributional
action contribution $\ud\hlag$  must be constructed in terms of the
corresponding regular d-surface density scalar $\ud\olag$ according to the
prescription that is expressible in standard Dirac notation as
\be  \ud\hlag=\Vert \bg\Vert^{-1/2}\int\!\! \ud d\!\oS\,
\ud\olag\, \delta^{\rm n}[x-x\{\sigme\}]\, .\eqn{2.3}\fe

\subsection{Current, generalised vorticity, and stress-energy tensor}
\label{2-2}

In the kind of model under consideration, each supporting d-surface is
supposed to be endowed with its own independent internal field variables which
are allowed to couple with each other and with their derivatives in the
corresponding d-surface Lagrangian contribution $\ud\olag$, and which are also
allowed to couple into the Lagrangian contribution $\udi\olag$ on any of its
attached boundary (d--1) surfaces, though -- in order not to violate the
strong dynamic regularity condition -- they are not allowed to couple into
contributions of dimension (d--2) or lower. As well as involving its own
d-brane surface fields and those of any (d+1) brane to whose boundary it may
belong, each contribution $\ud\olag$ may also depend passively on the fields
of a fixed higher dimensional background. Such fields will of course always
include the background spacetime metric $\bg_{\mu\nu}$ itself. Apart from that,
the most commonly relevant kind of backround field (the only one allowed for
in the earlier analysis, \cite{Carter90}) is a Maxwellian gauge potential $\bA_\mu$ whose
exterior derivative is the automatically ``closed" electromagnetic field,
\be \bF_{\mu\nu}=2\nabl_{[\mu}\bA_{\nu]} \ , \hskip 1 cm
\nabl_{[\mu}\bF_{\nu\rho]}= 0\, .\eqn{2.4}\fe

Although many other possibilities can in principle be envisaged, the most 
commonly relevant generalisation, to which for the sake of simplicity the  
following analysis will be limited, consists of allowance  just for
another background field of the generic Ramond type (of which the ordinary 
gauge covector $\bA_\nu$ is a special single index case) that is 
important in wide range of applications including the kind of cosmic or
superfluid defects for which this work is particularly intended,
namely a gauge  r - form, i.e.  an antisymmetric covariant r index tensor 
field with components $\bB_{\mu\nu...}=\bB_{[\nu\mu...]}$,  
whose exterior derivative is an automatically closed physical current
(r+1) - form,
\be \bN_{\mu\nu\rho...}=({\rm r}+1)\nabl_{[\mu}\bB_{\nu\rho...]} \, ,
\hskip 1 cm
\nabl_{[\mu}\bN_{\nu\rho\sigma...]}=0\, .\eqn{2.5}\fe
Just as a Maxwellian gauge transformation of the form $\bA_\mu\mapsto
\bA_\mu+\nabl_\mu\alpha$ for an arbitrary scalar $\alpha$ leaves the
electromagnetic field (\ref{2.4}) invariant, so analogously a Kalb-Ramond 
gauge transformation $\bB_{\mu\nu...}\mapsto \bB_{\mu\nu...}+
{\rm r}!\nabl_{[\mu}\chi_{\nu...]}$ for an arbitrary (r--1) - form 
$\chi_{\mu...}$ leaves the corresponding current (r+1) - form 
(\ref{2.5}) invariant. 

An example of the kind that is most common in an ordinary 4 - dimensional 
spacetime background is that of a Kalb - Ramond field, meaning a 2 index 
Ramond field with components $\bA\!^{_{\{2\}}}{\!}_{\mu\nu}
=-\bA\!^{_{\{2\}}}{\!}_{\nu\mu}$ for which the corresponding
current 3 form $\bF\!^{_{\,\{3\}}}{\!}_{\mu\nu\rho}=$ $\nabl_\mu 
\bA\!^{_{\{2\}}}{\!}_{\nu\rho}+\nabl_\nu \bA\!^{_{\{2\}}}{\!}_{\rho\mu}+
\nabl_\rho \bA\!^{_{\{2\}}}{\!}_{\mu\nu}$  will just be the dual
$\bF\!^{_{\,\{3\}}}{\!}_{\mu\nu\rho}=\bepsilon_{\mu\nu\rho\sigma} n^\sigma$ of 
an ordinary current vector  $n^\mu$ satisfying a conservation law of the 
usual type, $\nabl_\mu n^\mu=0$. Such a Kalb-Ramond representation can be 
used to provide an elegant variational formulation for ordinary perfect 
fluid theory \cite{Carter94} and is particularly convenient for setting 
up ``global" string models of vortices both in a simple cosmic axion or 
Higgs field \cite{VilenkinVachaspati87,DavisShellard89,Sakellariadou91} 
and in a superfluid \cite{BenYaacov92} such as liquid Helium-4.

In accordance with the preceeding considerations, the analysis that follows
will be based on the postulate that the action is covariantly and gauge
invariantly determined by specifying each scalar Lagrangian contribution
$\ud\olag$ as a function just of the background fields, $\bA_\mu$,
$\bB_{\mu\nu...}$ and of course $\bg_{\mu\nu}$, and of any relevant internal 
fields (which in the simplest non-trivial case -- exemplified by  string
models \cite{Carter89,Larsen93} of the category needed for the 
macroscopic description of Witten type \cite{Witten85} superconducting 
vortices -- consist just of a phase scalar $\phie$). In accordance with the 
restriction that the branes be ``pure" or ``fine" in the sense explained 
above, it is postulated that polarisation effects are excluded by ruling out 
couplings involving gradients of the background fields. This means that the 
effect of making arbitrary infinitesimal ``Lagrangian" variations 
$\dL \bA_\mu$, $\dL \bB_{\mu\nu...}$, $\dL \bg_{\mu\nu}$ of the background 
fields will be to induce a corresponding variation $\d\Ac$ of the action that 
simply has the form
$$\d \Ac=\sum_{\rm d}\int\!\! \ud d\!\oS\Big\{
\ud\oj{^\mu} \dL \bA_\mu +{1\over {\rm r}!}\ud\oW{^{\mu\nu...}}
\dL \bB_{\mu\nu...}\hskip 1 cm $$ \be\hskip 1 cm + 
{_1\over^2} \ud\oT{^{\mu\nu}} \dL  \bg_{\mu\nu} \Big\} \, , \eqn{2.6}\fe
provided either that that the relevant independent internal field components
are fixed or else that the internal dynamic equations of motion are
satisfied in accordance with the variational principle stipulating that
variations of the relevant independent field variables should make no
difference. For each d-brane of the complex, as well as the 
{ surface stress momentum energy density} tensor 
$\ud\oT{^{\mu\nu}}=\ud\oT{^{\nu\mu}}$,
this partial differentiation formula also 
implicitly specifies the corresponding {electromagnetic surface current 
density} vector $\ud\oj{^{\mu}}$, and the (antisymmetric)  { surface 
flux} r - vector $\ud \oW{^{\mu\nu...}}=\ud\oW{^{[\mu\nu...]}}$,  which
is interpretable as vorticity in the 2-index Kalb Ramond case.  
These quantities are formally expressible more explicitly as 
\be \ud\oj{^{\mu}} = {\partial\!\ud \olag\over\partial \bA_\mu }
\, , \hskip 0.6 cm 
\ud\oW{^{\mu\nu...}} = {\rm r}!{\partial\! \ud\olag\over\partial 
\bB_{\mu\nu...}}\, ,\fe 
and 
\be \ud\oT{^{\mu\nu}}=2{\partial\!\ud \olag\over\partial 
\bg_{\mu\nu}}+\ud\olag\ud\eta^{\mu\nu}  
\, , \eqn{2.7}\fe
of which the latter is obtained using the formula
\be \dL (\ud d\!\oS)={_1\over^2}\ud\og^{\mu\nu}
(\dL \bg_{\mu\nu}) \ud d\!\oS\, ,\eqn{varmes}\fe
where $\ud\og^{\mu\nu}$ is the rank - d  {fundamental tensor} of the
d - dimensional imbedding, as defined in the manner described in the
preceeding section.

\subsection{Conservation of current and generalised vorticity}
\label{2-3}

The condition that the action be gauge invariant means that if one simply
sets $\dL \bA_\mu=\nabl_\mu \alpha$, $\dL \bB_{\mu\nu...}
={\rm r}!\nabl_{[\mu}\chi_{\nu...]}$, $d_{_{\rm L}}\bg_{\mu\nu}=0$, 
for arbitrarily chosen $\alpha$ and $\chi_{\mu...}$ then $\d\Ac$ should 
simply vanish, i.e.
\be \sum_{\rm d}\int\!\! d\!\ud\oS\left\{
\ud\oj{^\mu} \nabl_\mu\alpha+\ud\oW{^{\mu\nu...}}\nabl_\mu\chi_{\nu...}
 \right\}=0 \, . \eqn{2.8}\fe
In order for this to be able to hold for an arbitrary scalar field $\alpha$ and an
an arbitrary (r--1) form $\chi_\mu$ it is evident that the surface current 
$\ud\oj{^\mu}$ and the (generalised vorticity)
 flux r - vector $\ud\oW{^{\mu\nu...}}$ must (as one would anyway expect
from the consideration that they depend just on the relevant internal
d-surface fields) be purely d-surface tangential, i.e. their contractions
with the relevant rank (n--d) orthogonal projector $\ud\!\!\ag^\mu_{\
\nu}=\bg^\mu_{\ \nu}- \ud\og^\mu_{\ \nu}$ must vanish:
\be \ud\!\!\ag^\mu_{\ \nu}\ud\oj{^\nu}=0\ , \hskip 1 cm \ud\!\!
\ag^\mu_{\ \nu}\ud\oW{^{\nu\rho...}}=0\, .\eqn{2.9}\fe
Hence, decomposing the full gradient operator $\nabl_\mu$ as the sum of
its tangentially projected part $\ud\onab_\mu=\ud\og^\nu_{\ \mu}\nabl_\nu$
and of its orthogonally projected part $\ud\!\!\ag^\nu_{\ \mu}\nabl_\mu$, 
and noting that by (\ref{2.9}) the latter will give no contribution, one sees 
that (\ref{2.8}) will take the form
\begin{eqnarray}
 \sum_{\rm d}\int\!\!\ud d\!\oS\Big\{ \ud\onab_\mu
\Big(\ud\oj{^\mu} \alpha+\ud\oW{^{\mu\nu...}}\chi_{\nu...} \Big)
\qquad & & \nonumber\\  
-\alpha\ud\onab_\mu\oj{^\mu} -\chi_{\nu...}\ud\onab_\mu
\ud\oW{^{\mu\nu...}}\Big\} & =0  \, , & \eqn{2.10}\end{eqnarray}
in which first term of each integrand is a pure surface divergence.
Such a divergence can be dealt with using Green's theorem, according to which,
for any d-dimensional support surface $\ud\oS$ of a (d--1)-brane,
one has  the identity
\be \int\!\! \ud d\!\oS\ud\onab_\mu\ud\oj{^\mu}=
\oint\! \udi d\!\oS\udp\lamb_\mu\ud\oj{^\mu}\, ,\eqn{2.11}\fe
where the integral on the right is taken over the boundary (d--1)-surface
$\partial\!\ud\oS$ of $\ud\oS$, and $\udp\lamb_\mu$ is the
(uniquely defined) unit tangent vector on the d-surface that is directed
normally outwards at its (d--1)-dimensional boundary. Bearing in mind that a
membrane support hypersurface can belong to the boundary of two distinct
media, and that for d$\leq$ n--3 a d-brane may belong to a common boundary
joining three or more distinct (d+1)-branes of the complex under
consideration, one sees that (\ref{2.10}) is equivalent to the condition
\begin{eqnarray} 
\sum_{\rm p}\int\!\! \up d\!\oS\,\Big\{ \alpha\Big(\!
\up\onab_\mu\up\oj{^\mu}\! -\sum_{\rm d=p+1}\!\!\udp\lamb_\mu
\ud\oj{^\mu}\Big) \qquad & & \nonumber\\ 
+ \chi_{\nu...}\Big(\!\up\onab_\mu\up\oW{^{\mu\nu...}}\! 
-\sum_{\rm d=p+1}\!\!\udp\lamb_\mu\ud\oW{^{\mu\nu...}}\Big)\Big\}
 & =0\, , & 
 \eqn{2.12}
\end{eqnarray}
where, for a particular p-dimensionally supported (p--1)-brane, the summation
``over d=p+1" is to be understood as consisting of a contribution from each
(p+1)-dimensionally supported p-brane attached to it, where for each such
p-brane, $\udp\lamb_\mu$ denotes the (uniquely defined) unit tangent vector
on its (p+1)-dimensional support surface that is directed normally towards the
p-dimensional support surface of the boundary (p--1)-brane. The Maxwell gauge
invariance requirement to the effect that (\ref{2.12}) should hold for
arbitrary $\alpha$ can be seen to entail an electromagnetic charge
conservation law of the form
\be \up\onab_\mu\up\oj{^\mu}=\sum_{\rm d=p+1}\!\!\udp\lamb_\mu
\ud\oj{^\mu}\, . \eqn{2.13}\fe
This can be seen from (\ref{2.11}) to be be interpretable as meaning
that the total charge flowing out of particular (d--1)-brane from its 
boundary is balanced by the total charge  flowing  into it from any d-branes 
to which it may be attached. The analogous Ramond gauge invariance 
requirement that (\ref{2.12}) should also hold for an arbitrary (r--1) - form
$\chi_{\mu...}$ can be seen to entail a corresponding (vorticity) flux
conservation law of the form
\be \up\onab_\mu\up\oW{^{\mu\nu...}}=\sum_{\rm d=p+1}\!\!
\udp\lamb_\mu\ud\oW{^{\mu\nu...}}\, .\eqn{2.14}\fe
A more sophisticated but less practical way of deriving the foregoing
conservation laws would be to work not from the expression (\ref{2.1}) in
terms  of ordinary surface integrals but instead to use the superficially
simpler expression (\ref{2.2}) in terms of distributions, which leads to the
replacement of (\ref{2.13}) by the ultimately equivalent (more formally
obvious but less directly meaningful) expression
\be \nabl_\mu\Big(\sum_{\rm d}\!\!\ud\hj{^{\mu}}\Big)=0 \eqn{2.15}\fe
involving the no longer regular but Dirac
distributional current $\ud\hj{^\mu}$ that is given in terms of the
corresponding regular surface current $\ud\oj{^\mu}$ by
\be \ud \hj{^{\mu}}=\Vert\bg \Vert^{-1/2}\int\!\!\ud d\!\oS\,
\ud \oj{^{\mu}}\, \delta^{\rm n}[x-x\{\sigme\}]\, .\eqn{2.16}\fe
Similarly one can if one wishes rewrite the flux conservation 
law (\ref{2.14}) in the distributional form
\be \nabl_\mu\Big(\sum\!\! \ud \hW{^{\mu\nu...}}\Big)=0\, , \eqn{2.17}\fe
where the distributional (generalised vorticity) flux 
$\ud \hW{^{\mu\nu...}}$ is given in terms of the corresponding regular 
surface flux $\ud\oW{^{\mu\nu...}}$ by
\be \ud \hW{^{\mu\nu...}}=\Vert\bg \Vert^{-1/2}\int\!\!\ud d\!\oS\,
\ud \oW{^{\mu\nu...}}\ \delta^{\rm n}[x-x\{\sigme\}]\, .\eqn{2.18}\fe
It is left as an entirely optional exercise for any  readers who may be adept
in distribution theory to show how the ordinary functional relationships
(\ref{2.13}) and (\ref{2.14}) can be recovered by by integrating out the 
Dirac distributions in (\ref{2.15}) and (\ref{2.17}).

\subsection{Force and the stress balance equation}
\label{2-4}

The condition that the hypothetical variations introduced in (\ref{2.6})
should be ``Lagrangian" simply means that they are to be understood to be
measured with respect to a reference system that is comoving with the various
branes under consideration, so that their localisation with respect to it
remains fixed. This condition is necessary for the variation to be meaningly
definable at all for a field whose support is confined to a particular brane
locus, but in the case of an unrestricted background field one can enviseage
the alternative possibility of an ``Eulerian" variation, meaning one defined
with respect to a reference system that is fixed in advance, independently of
the localisation of the brane complex, the standard example being that of a
Minkowski reference system in the case of a background that is flat. In such a
case the relation between the  more generally meaningfull Lagrangian
(comoving) variation, denoted by $\dL $, and the corresponding
Eulerian (fixed point) variation denoted by $\dE $ say will be given
by Lie differentiation with respect to the vector field $\xi^\mu$ say that
specifies the infinitesimal of the comoving reference system with respect to
the fixed background, i.e. one has
\be \dL -\dE=\vec{\ \xi\Libra}\, ,\eqn{2.19}\fe
where the Lie differentiation operator $\vec{\ \xi\Libra}$ is given for the
background fields under consideration here by
\be \vec{\ \xi\Libra} \bA_\mu = \xi^\sigma\nabl_\sigma 
\bA_\mu+\bA_\sigma\nabl_\mu\xi^\sigma \, , \fe \be
\vec{\ \xi\Libra} \bB_{\mu\nu...}=\xi^\sigma\nabl_\sigma \bB_{\mu\nu}
+{\rm r}!\bB_{\sigma[\nu...}\nabl_{\mu]}\xi^\sigma \, , \fe \be
\vec{\ \xi\Libra} \bg_{\mu\nu}= 2\nabl_{(\mu}\xi_{\nu)}\, .\eqn{2.20}\fe
 
This brings us to the main point of this section which is the derivation of
the dynamic equations governing the extrinsic motion of the branes of the
complex, which are obtained from the variational principle to the effect that
the action $\Ac$ is left invariant not only by infinitesimal variations of the
relevant independent intrinsic fields on the support surfaces but also by
infinitesimal displacements of the support surfaces themselves. Since the background 
fields $\bA_\mu$, $\bB_{\mu\nu...}$, and $\bg_{\mu\nu}$ are to be considered
as fixed, the relevant Eulerian variations simply vanish, and so the resulting
Lagrangian variations will be directly identifiable with the corresponding Lie
derivatives -- as given by (\ref{2.20}) -- with respect to the generating
vector field $\xi^\mu$ of the infinitesimal displacement under consideration.
The variational principle governing the equations of extrinsic motion is thus
obtained by setting to zero the result of substituting these Lie derivatives
in place of the corresponding Lagrangian variations in the more general
variation formula (\ref{2.6}), which gives
\begin{eqnarray} \sum_{\rm d}\int\!\! \ud d\!\oS\Big\{
\ud\oj{^\mu} \vec{\ \xi\Libra} \bA_\mu +{1\over {\rm r}!}\ud\oW{^{\mu\nu...}}
\vec{\ \xi\Libra} \bB_{\mu\nu...} \qquad & & \nonumber\\ 
+ {_1\over^2} \ud\oT{^{\mu\nu}}\vec{\ \xi\Libra}
\bg_{\mu\nu} \Big\} &=0 \, . & \eqn{2.21}\end{eqnarray}
The requirement that this should hold for any choice of $\xi^\mu$ evidently
implies that the tangentiality conditions (\ref{2.9}) for the surface fluxes
$\ud\oj{^\mu}$ and $\ud\oW{^{\mu\nu}}$ must be supplemented by an
analogous d-surface tangentiality condition for the surface stress momentum
energy tensor $\ud\oT{^{\mu\nu}}$, which must satisfy
\be \ud\!\!\ag^\mu_{\ \nu}\ud\oT{^{\nu\rho}}=0\, .\eqn{2.22}\fe
(as again one would  expect anyway from the consideration that it depends just
on the relevant internal d-surface fields). This allows  (\ref{2.20}) to be
written out in the form
\begin{eqnarray} 
\sum_{\rm d}\int\!\!\ud d\!\oS\Big\{
\xi^\rho\Big(\bF_{\rho\mu}\ud\oj{^\mu}\!+{1\over {\rm r}!}\bN_{\rho\mu\nu...}
\ud\oW^{\mu\nu...}\! \qquad\qquad & & \nonumber\\ 
-\ud\onab_\mu\ud\oT{^\mu_\rho}-\bA_\rho\ud\onab_\mu \ud\oj{^\mu}\! 
-{1\over({\rm r}-1)!}\bB_{\rho\nu...}\ud\onab_\mu\ud\oW{^{\mu\nu...}}\Big) 
& & \nonumber\\  
+ \ud\onab_\mu\Big(\xi^\rho( \bA_\rho \ud\oj{^\mu}\! 
+{1\over({\rm r}-1)!} \bB_{\rho\nu...} \ud\oW{^{\mu\nu...}}\!+
\ud\oT{^\mu}_\rho)\Big) \Big\}=0  \, , & & \eqn{2.23}
\end{eqnarray}
in which the final contribution is a pure surface divergence that can be dealt
with using Green's theorem as before. Using the results (\ref{2.13}) and
(\ref{2.14}) of the analysis of the consequences of gauge invariance and
proceeding as in their derivation above, one sees that the condition for
(\ref{2.23}) to hold for an arbitrary field $\xi^\mu$ is that, on each
(p--1)-brane of the complex, the dynamical equations
\be \up\onab_\mu\up\oT{^\mu}_\rho= \up \tf_\rho \, , \eqn{2.24}\fe
should be satisfied for a total force density $\up \tf_\rho$ given by
\be \up \tf_\rho=\up\of_\rho+\up\cf_\rho \, , \eqn{2.24a}\fe
where $\up\cf_\rho$ is the contribution of the contact force exerted
on the p-surface by other members of the brane complex, which takes the
form
\be \up\cf_\rho= \sum_{\rm d=p+1}\!\!
\udp\lamb_\mu\ud\oT{^\mu}_\rho  \, ,\eqn{2.25}\fe
while the other force density contribution $\up\of_\rho$ represents
the effect of the external background fields, which  is given by
\be \up\of_\rho=\bF_{\rho\mu}\up\oj{^\mu}+{1\over {\rm r}!}
\bN_{\rho\mu\nu...}\up\oW{^{\mu\nu...}}\, .\eqn{2.26}\fe
As before, the summation ``over d=p+1" in (\ref{2.25}) is to be understood as
consisting of a contribution from each of the p-branes attached to the (p--1)-
brane under consideration, where for each such attached p-brane,
$\udp\lamb_\mu$ denotes the (uniquely defined) unit tangent vector on its
(p+1)-dimensional support surface that is directed normally towards the
p-dimensional support surface of the boundary (p--1)-brane. 

The first of the background force contibutions in (\ref{2.26}) is of 
course the Lorentz type force density resulting from the effect of the 
electromagnetic field on the surface current. 
For the case of an ordinary current 3-vector $\bN_{\rho\mu\nu}$,
the other contribution in (\ref{2.26}) will just be the Joukowsky
type force density (of the kind responsible for the lift on an aerofoil)
resulting from the Magnus effect, which acts in the case of a ``global"
string \cite{VilenkinVachaspati87,DavisShellard89} though not in the case of a
string of the ``local" type for which the relevant vorticity flux
$\up\oW{^{\mu\nu}}$ will be zero. As with the conservation laws (\ref{2.13})
and (\ref{2.14}), so also the explicit force density balance law expressed by
(\ref{2.24})  can alternatively be expressed in terms of the corresponding
Dirac distributional stress momentum energy and background force density
tensors, $\ud\hT^{\mu\nu}$ and $\ud\hf_\mu$, which are given for each
(d--1)-brane in terms of the corresponding regular surface stress momentum
energy and background force density tensors $\ud\oT{^{\mu\nu}}$ and
$\ud\of_\mu$ by 
\be  \ud \hT^{\mu\nu}=\Vert\bg \Vert^{-1/2}\int\!\!\ud d\!\oS\,
\ud \oT{^{\mu\nu}}\, \delta^{\rm n}[x-x\{\sigme\}]\ \eqn{2.27}\fe
and
\be \ud \hf_\mu=\Vert\bg \Vert^{-1/2}\int\!\!\ud d\!\oS\,
\ud \of_\mu \ \delta^{\rm n}[x-x\{\sigme\}]\, . \eqn{2.28}\fe
The equivalent -- more formally obvious but less explicitly meaningful --
distributional versional version of the force balance law (\ref{2.24}) takes
the form
\be \nabl_\mu\Big(\sum_{\rm d}\!\! \ud \hT{^{\mu}_{\ \rho}}\Big)
=\hf_\rho\, , \eqn{2.29}\fe
where the total Dirac distributional force density is given in terms of the 
electromagnetic current distributions (\ref{2.16}) and the (generalised vorticity)
flux distributions (\ref{2.18}) by
\be \hf_\rho=\bF_{\rho\mu}\sum_{\rm d}\!\!\ud\hj{^\mu}+{1\over {\rm r}!}
\bN_{\rho\mu\nu...}\sum_{\rm d}\!\!\ud\hW{^{\mu\nu...}}\, .\eqn{2.30}\fe
It is again left as an optional exercise for readers who are adept in the use
of Dirac distributions to show that the system (\ref{2.24}), (\ref{2.25}),
(\ref{2.26}) is obtainable from (\ref{2.29}) and (\ref{2.30}) by substituting
(\ref{2.16}), (\ref{2.18}), (\ref{2.27}), (\ref{2.28}).

As an immediate corollary of (\ref{2.24}), it is to be noted that for any 
vector field $\el^\mu$ that generates a continuous symmetry of the background
spacetime metric, i.e. for any solution of the Killing equations
\be \nabl_{(\mu}\el_{\nu)}=0\, ,\eqn{2.31}\fe
one can construct a corresponding surface momentum or energy density current
\be \up\oP{^\mu}=\up\oT{^{\mu\nu}}\el_\nu\ ,\eqn{2.32}\fe
that will satisfy
\be \up\onab_\mu\up\oP{^\mu}= \sum_{\rm d=p+1}\!\!
\udp\lamb_\mu\ud\oP{^\mu}+\up\of_\mu k^\mu  \, .\eqn{2.33}\fe
In typical applications for which the n-dimensional background spacetime can
be taken to be flat there will be n independent translation Killing vectors
which alone (without recourse to the further n(n--1)/2 rotation and boost
Killing vectors of the Lorentz algebra) will provide a set of relations of the
form (\ref{2.33}) that together provide the same information as that in the
full force balance equation (\ref{2.24}) or (\ref{2.29}).

\subsection{The equation of extrinsic motion}
\label{2-5}

Rather than the distributional version (\ref{2.29}), it is the explicit
version (\ref{2.24}) of the force balance law that is directly useful for
calculating the dynamic evolution of the brane support surfaces. Since the
relation (\ref{2.29}) involves n independent components whereas the support
surface involved is only p-dimensional, there is a certain redundancy, which
results from the fact that if the virtual displacement field $\xi^\mu$ is
tangential to the surface in question it cannot affect the action. Thus if
$\up\!\!\ag^{\!\mu}_{\,\nu}\xi^\nu=0$, the condition (\ref{2.21}) will be
satisfied as a mere identity -- provided of course that the field equations
governing the internal fields of the system are satisfied. It follows that the
non-redundent information governing the extrinsic motion of the p-dimensional
support surface will be given just by the orthogonally projected part of
(\ref{2.24}). Integrating by parts, using the fact that, by (\ref{1.6}) and
(\ref{1.15}), the surface gradient of the rank-(n--p) orthogonal projector
$\up\!\!\ag^{\!\mu}_{\,\nu}$ will be given in terms of the second
fundamental tensor $\up \Ke_{\mu\nu}^{\ \ \,\rho}$ of the p-surface by
\be \up\onab_\mu\up\!\!\ag^{\!\nu}_{\,\rho}=-\up \Ke_{\mu\nu}^{\ \ \,\rho}
-\up \Ke_{\mu\ \nu}^{\ \rho} \, ,\eqn{2.34}\fe
it can be seen that the extrinsic equations of motion obtained as the
orthogonally projected part of (\ref{2.24}) will finally be expressible by
\be \up\oT{^{\mu\nu}}\up \Ke_{\mu\nu}^{\ \ \,\rho}=
\up\!\!\ag^{\!\rho}_{\,\mu}\up \tf^\mu\, .\eqn{2.35}\fe

It is to be emphasised that the formal validity of the formula that has just
been derived is not confined to the variational models on which the above
derivation is based, but also extends to dissipative models (involving effects
such as external drag by the background medium 
\cite{Carteretal94,Vilenkin91,GarrigaSakellariadou93} or mutual resistance
between independent internal currents). The condition that even a
non-conservative macroscopic model should be compatible with an underlying
microscopic model of conservative type  requires the existence (representing
to averages of corresponding microscopic quantities) of appropriate stress
momentum energy density and force density fields satisfying (\ref{2.35}). 

The ubiquitously applicable formula (\ref{2.35})is interpretable as
being just the natural higher generalisation of ``Newton's law"
(equating the product of mass with acceleration to the applied force)
in the case of a particle. The surface stress momentum energy tensor,
$\up\oT{^{\mu\nu}}$, generalises the mass, and the second fundamental
tensor, $\up \Ke_{\mu\nu}^{\ \ \,\rho}$, generalises the acceleration.

The way this works out in the 1-dimensional case of a ``pure" point
particle (i.e. a monopole) of mass $\mag$, for which the Lagrangian is
given simply by $^{_{(1)}}\!\olag=-\mag$, is as follows. The 1-dimensional
energy tensor will be obtained in terms of the unit tangent vector
$\ue^\mu$ ($\ue^\mu \ue_\mu=-1$) as $^{_{(1)}}\! \oT{^{\mu\nu}}=\mag\, \ue^\mu
\ue^\nu$, and in this zero-brane case, the first fundamental tensor will
simply be given by $^{_{(1)}}\!\og^{\mu\nu}=-\ue^\mu \ue^\nu$, so that the
second fundamental tensor will be obtained in terms of the acceleration
$\dot \ue^\mu =\ue^\nu\nabl_\nu \ue^\mu$ as
$^{_{(1)}}\!\Ke_{\mu\nu}^{\ \ \,\rho} =\ue_\mu \ue_\nu \dot \ue^\rho$. Thus
(\ref{2.35}) can be seen to reduce in the case of a particle simply to
the usual familiar form $\mag\, \dot \ue^\rho=
^{_{(1)}}\!\ag^{\!\rho}_{\,\mu}{^{_{(1)}}}\! \tf^\mu$.

The familiar electromagnetic example of the Faraday - Lorentz force exerted
on a charged point particle (i.e. a zero brane) by an ordinary Maxwellian  
field is the simplest example of the effect of the important special case of 
what (in view of the proverbial complementarity of ``brain versus brawn'')
may conveniently be termed the relevant ``brawn field''. For
a generic (p--1) brane, with worldsheet dimension p,  the corresponding
brawn field is defined to be a Ramond type gauge r form
$\bB_{\mu\nu...}$ whose index number r is equal to the worldsheet
dimension, i.e. for which r=p. In this case the
 corresponding
generalised vorticity flux on the brane must evidently be given by
an expression of the form
\be \oj\!_{_{\{\rm p\}}}\!{^{\mu\nu...}} =e_{_{\{\rm p\}}}\ \up
\calE^{\mu\nu ...}
\, ,\eqn{2.50}\fe
for some proportionality factor $e_{_{\{\rm p\}}}$. Moreover, provided
that this  brawn source flux is confined to the  d dimensional
brane worldsheet, so that the right hand side of the flux conservation
law (\ref{2.14}) vanishes, this proportionality factor must have
vanishing worldsheet gradient,
\be \up\onab_\nu e_{_{\{\rm p\}}}=0\, ,\eqn{2.51}\fe
 so that
$e_{_{\{\rm p\}}}$ will have a fixed value. The coefficient 
$e_{\{{\rm p\}}}$ will thus be interpretable as a 
brawn charge coupling constant characterising the p-brane.
In particular, for the case of a zero brane (i.e. a point particle)
the relevant coupling constant $e_{_{\{1\}}}$ will be interpretable as 
an ordinary electromagnetic charge. Similarly for a 1-brane
(i.e. a string) the relevant (Wess-Zumino type) coupling constant
$e_{_{\{2\}}}$ will be interpretable as a measure of the relevant Kalb-Ramond
current circulation round the worldsheet. When the relevant ``brawn''
field provides the only external force on the brane the orthogonal
projection on the right of (\ref{2.35}) will be redundant, and
the equation of extrinsic motion of the worldsheet will reduce to the
explicit form
\be \up\oT{^{\mu\nu}}\up \Ke_{\mu\nu\rho}= {e_{_{\{\rm p\}}}
\over {\rm p}!}\, \bF^{_{\{\rm d\}}}\!_{\rho\sigma ...}\ \up
\calE^{\sigma ...}\, ,
\eqn{2.52}\fe
with d=p+1 as before. 
For the case of a point particle in an electromagnetic field this is just the
usual equation of motion provided by the Faraday Lorentz force, while for the
case of a string surrounded by a Kalb-Ramond current this is just the equation
of motion provided~\cite{CarterLanglois95} by the Joukowski lift force density
that is attributable to the familiar Magnus effect. For the 
source free dynamical equation governing the ``brawn field'' outside the 
brane, the simplest possibility is a divergence equation of the 
familiar form
\be \nabla^\rho  \bF^{_{\{\rm d\}}}\!_{\rho\sigma ...}
=0\, ,\eqn{2.53}\fe
which applies both to ordinary Maxwellian electromagnetism and  to the
standard kind of axion fluid model~
\cite{DabholkarQuashnock90,Carter94,BattyeShellard95,BattyeShellard96}

The possibility that such an effect occurs for the 3-brane of a ``brane 
world'' scenario has not yet received much attention, presumably because a 
non-zero value for the relevant generalised Wess-Zumino coupling constant
$e_{_{\{4\}}}$ and the ensuing specification of a preferred orientation 
in the worldsheet (due to the pseudo-tensorial, not strictly tensorial, nature 
of the 4 - surface alternating tensor $\calE^{\mu\nu\rho\sigma}$) would have 
no effect in a scenario respecting the reflection invariance of the
5-dimensional scenarios that are most commonly considered
\cite{ChamGib99,ChamEtal99,BinEtal99,ShirEtal00,GregEtal00,Ms00,MLW00,BatMen00}.
However the consequences of dropping the $Z_2$ reflection symmetry constraint have 
recently begun to be a subject of systematic 
investigation~\cite{STW00,Gregetal00,KMPRS99,KMPR00,DDPV00}. In the absence of 
reflection symmetry a generalised Wess-Zumino type coupling effect of the type 
characterised by (\ref{2.52}) can provide a plausible underlying mechanism that, 
subjet to (\ref{2.53}), would simulate a phase transiion of the cosmological 
``constant'' of the ``bulk'',  such as has recently been postulated in bubble type 
scenarios~\cite{DerDol00,Per00} of this less orthodox $Z_2$ symmetry 
violating kind.

For a brane of codimension 1,  i.e in a background of dimension  n=d=p+1, the 
external ``brawn'' field $\bF^{_{\{\rm d+1\}}}\!_{\rho\sigma ...}$ must 
evidently be proportional to the background measure tensor 
$\varepsilon_{\rho\sigma ...}$, with a proportionality factor that must be 
uniform over any region where the source free field equation (\ref{2.53}) is 
satisfied, so that for a (d--1) - brane in a (d+1) dimensional bulk we shall 
have $\bF^{_{\{\rm d+1\}}}\!_{\mu\nu ...}= \bF^{_{\{\rm d+1\}}} \,
\varepsilon_{\mu\nu ...}$ with a ``brawn'' field pseudoscalar
$\bF^{_{\{\rm d+1\}}}$ that has constant value (giving a stress energy density 
tensor of the same form as would arise from a cosmological constant 
proportional to $|\bF|^2$)  which will give rise to a force density with 
uniform magnitude proportional to the product $e_{_{\{\rm d\}}}
\bF^{_{\{\rm d+1\}}}$. Thus using the unit normal
$\lambda_\mu=({\rm d}!)^{-1}\varepsilon_{\mu\nu\ ...}
{\,^{\,_{(\rm d)}}}\!\calE^{\nu ...}$ (with d=4 in the usual brane world 
case) to construct the (symmetric) second fundamental form
${\,^{\,_{({\rm d})}}}\!\Ke_{\mu\nu}={\,^{\,_{({\rm d})}}}\!
\Ke_{\mu\nu}^{\ \ \,\rho}\lambda_\rho$,  it can be seen that the equation
of motion (\ref{2.52}) will be expressible in this case as
\be {\,^{\,_{({\rm d})}}}\!\oT{^{\mu\nu}}{\,^{\,_{({\rm d})}}}\! \Ke_{\mu\nu}
=e_{_{\{{\rm d}\}}}\bF^{_{\{\rm d+1\}}}\, .\eqn{2.55}\fe
(In a more elaborate treatment allowing for the active role of the
brane as a source for the ``brawn'' field, this constant product
$e_{_{\{{\rm d}\}}}\bF^{_{\{\rm d+1\}}}$ would need to be replaced by a 
constant proportional to the resulting surface discontinuity in 
$\big(\bF^{_{\{\rm d+1\}}}\big)^2$, and if self gravitation were also taken 
into account then (as will be discussed in more detail elsewhere) the tensor 
${\,^{\,_{({\rm d})}}}\! \Ke_{\mu\nu}$ would also be continuous and its value 
in (\ref{2.55}) would need to be replaced by the mean of its values on the 
two sides.)

\subsection{Perturbations and extrinsic characteristic equation} 
\label{3-1}
 
Two of the most useful formulae for the analysis of small perturbations of a
string or higher brane worldsheet are the expressions for the infinitesimal
Lagrangian (comoving) variation of the first and second fundamental tensors in
terms of the corresponding comoving variation $\dL \bg_{\mu\nu}$ of the metric
(with respect to the comoving reference system). For the first fundamental
tensor one easily obtains 
\be\dL\og^{\mu\nu}= -\og^{\mu\rho}\og^{\mu\sigma}\dL \bg_{\rho\sigma}\ ,
\hskip 1 cm\dL \og^\mu_{\ \nu}=\og^{\mu\rho}\!\ag^{\!\sigma}_{\,\nu}
\dL \bg_{\rho\sigma} 
\eqn{3.1}\fe
and, by substituting this in the defining relation (\ref{1.10}),  the
corresponding Lagrangian variation of the second fundamental tensor is
obtained \cite{Carter93} as 
\be \dL \Ke_{\mu\nu}^{\ \ \,\rho}=\ag^{\!\rho}_{\,\lambda}\og^\sigma_{\,\mu}
\og^\tau_{\,\nu}\,\dL\Gamma_{\sigma\ \tau}^{\ \lambda} +\big(2\ag^{\!\sigma}
_{\,(\mu} \Ke_{\nu)}^{\ \ \tau\rho}-\Ke_{\mu\nu}^{\ \ \,\sigma}\og^{\tau\rho}
\big)\dL \bg_{\sigma\tau} \, ,\eqn{3.2}\fe
where the Lagrangian variation of the connection (\ref{1.18}) is given 
by the well known formula
\be \dL\Gamma_{\sigma\ \tau}^{\ \lambda} =g^{\lambda\rho}\big(
\nabl_{(\sigma\,}\dL \bg_{\tau)\rho}-{_1\over^2}\nabl_{\rho\,}\dL \bg_{\sigma\tau}
\big)\, .\eqn{3.3}\fe
Since we are concerned here only with cases for which the background is fixed
in advance so that the Eulerian variation $d_{\rm_E}$ will vanish in
(\ref{2.19}), the Lagrangian variation of the metric will be given just by its
Lie derivative with respect to the infinitesimal displacement vector field
$\xi^\mu$ that generates the displacement of the worldsheet under
consideration, i.e. we shall simply have
\be \dL \bg_{\sigma\tau}=2\nabl_{(\sigma}\xi_{\tau)} \, .\eqn{3.4}\fe
It then follows from (\ref{3.3}) that the Lagrangian variation of the
connection will be given by
\be \dL\Gamma_{\sigma\ \tau}^{\ \lambda}=\nabl_{(\sigma}\nabl_{\tau)}
\xi^\lambda-\calR^\lambda_{\ (\sigma\tau)\rho}\xi^\rho\, ,\eqn{3.5}\fe
where $\calR^\lambda_{\ \sigma\tau\rho}$ is the background Riemann curvature 
(which will be negligible in typical applications for which the lengthscales
characterising the geometric features of interest will be small compared
with those characterising any background spacetime curvature).
The Lagrangian variation of the first fundamental tensor is thus
finally obtained in the form
\be \dL\og^{\mu\nu}=-2\og_\sigma^{\,(\mu}\onab{^{\,\nu)}}\xi^\sigma\, ,
\eqn{3.6}\fe
while that of the second fundamental tensor is found to be given by
\begin{eqnarray}
 \dL \Ke_{\mu\nu}^{\ \ \,\rho}=\ag^{\!\rho}_{\,\lambda}\big(
\onab_{(\mu}\onab_{\nu)}\xi^\lambda-\og^\sigma_{\,(\mu}
\og^\tau_{\ \nu)}\calR^\lambda_{\ \sigma\tau\rho}\xi^\rho-\Ke^\sigma_{\ (\mu\nu)}
\onab_\sigma\xi^\lambda\big)+\nonumber\\ \hskip 2cm
\big(2\ag^{\!\sigma}_{\,(\mu}\Ke_{\nu)\tau}^{\ \ \ \rho}-g^\rho_{\,\tau}
\Ke_{\mu\nu}^{\ \ \,\sigma}\big)\big(\nabl_\sigma\xi^\tau +
\onab{^{\,\tau}}\xi_\sigma\big)\,  .\eqn{3.7}\end{eqnarray}

It is instructive to apply the forgoing formulae to the case of a
{ free} pure brane worldsheet, meaning one for which there is no external
force contribution so that the equation of extinsic motion reduces to the
form
\be \oT{^{\mu\nu}}\Ke_{\mu\nu}^{\ \ \,\rho}=0\, .\eqn{3.8}\fe
On varying the relation (\ref{3.8}) using (\ref{3.7}) in conjunction with the
orthogonality property (\ref{2.22}) and the unperturbed equation (\ref{3.8})
itself, the equation governing the propagation of the infinitesimal
displacement vector is obtained in the form
\be \ag^{\!\rho}_{\, \lambda}\oT{^{\mu\nu}}\big(\onab_\mu\onab_\nu
\xi^\lambda-\calR^\lambda_{\ \mu\nu\sigma}\xi^\sigma\big)
=-\Ke_{\mu\nu}^{\ \ \,\rho\,}\dL \oT{^{\mu\nu}}\ .\eqn{3.9}\fe

In the simplest case, for which there are no internal fields, (\ref{3.9})
constitutes the complete system of dynamical equations, which take an explicit
form~\cite{BattyeCarter95} that can be shown~\cite{BattyeCarter00} to be
directly obtainable by application of the variation principle to the second
order perturbation of the relevant Dirac-Goto-Nambu action. However, in the
generic case, the extrinsic perturbation equation (\ref{3.9}) will by itself
be only part of the complete system of perturbation equations governing the
evolution of the brane, the remaining equations of the system being those
governing the evolution of whatever surface current \cite{Carter89a} and other
relevant internal fields on the supporting worldsheet may be relevant. The
perturbations of such fields are involved in the source term on the right of
(\ref{3.9}), whose explicit evaluation depends on the specific form of the
relevant currents or other internal fields. However it is not necessary to
know the specific form of such internal fields for the purpose just of
deriving the characteristic velocities of propagation of the extrinsic
propagations represented by the displacement vector $\xi^\mu$, so long as they
contribute to the source term on the right of the linearised perturbation
equation (\ref{3.9}) only at first differential order, so that the
characteristic velocities will be completely determined by the first term on
the left of (\ref{3.9}) which will be the only second differential order
contribution. It is apparent from (\ref{3.9}) that under these conditions the
equation for the characteristic tangent covector $\chie_\mu$ say will be given
independently of any details of the surface currents or other internal fields
simply \cite{Carter90} by \be \oT{^{\mu\nu}}\chie_\mu\chie_\nu=0\,
.\eqn{3.10}\fe (It can be seen that the unperturbed surface stress momentum
energy density tensor $\oT{^{\mu\nu}}$ plays the same role here as that of the
unperturbed metric tensor $\bg^{\mu\nu}$ in the analogous characteristic
equation for the familiar case of a massless background spacetime field, as
exemplified by electromagnetic or gravitational radiation.)

\section{Acknowledgements}

I wish to thank B. Allen, C. Barrab\`es, R. Battye, A-C. Davis, 
R. Davis, V. Frolov, G. Gibbons, R. Gregory, T. Kibble, D. Langlois,
K. Maeda, X. Martin, P. Peter, T. Piran, D. Polarski, M. Sakellariadou, 
P. Shellard, P. Townsend, N. Turok, T. Vachaspati, and  A. Vilenkin, 
for many stimulating or clarifying discussions.


\begin{thebibliography}{000}

\bibitem{Carter95} B. Carter, 
``Dynamics of cosmic strings and other brane models'',
in {\it Formation and Interactions of Topological 
Defects} (NATO ASI {\bf B349}), ed. R. Brandenberger, A.-C. Davis,
 304-348 (Plenum, New York, 1995) [hep-th/9611054] 

\bibitem{Dirac62} P.A.M. Dirac, 
``An extensible model of the electron'',
{\it Proc. Roy. Soc. Lond.} {\bf A268},  57-67 (1962)

\bibitem{HoweTucker77} P.S. Howe, R.W. Tucker,
``A locally supersymmetric and reparametrisation invariant action for a
spinning membrane'',
 {\it J. Phys.} {\bf A10}, L155-62 (1977)

\bibitem{Achucarroetal87} A. Ach\'ucarro, J. Evans,  P.K. Townsend, 
D.L. Wiltshire,
``Super {\it p}-branes'',
{\it Phys. Lett.} {\bf 198 B}, 441-446 (1987)

\bibitem{Carter90}  B. Carter,  
``Covariant mechanics of simple and conducting cosmic strings and membranes'',
 in {\it Formation and Evolution of Cosmic Strings,} 
ed. G. Gibbons, S. Hawking, T. Vachaspati, {\it pp} 143-178 
(C.U.P., 1990)

\bibitem{Kibble76} T.W.B. Kibble, 
``Topology of cosmic domains and strings'',
{\it J. Phys.} {\bf A9},  1387-98 (1976)

\bibitem{VilenkinEverett82} A. Vilenkin, A.E. Everett, 
``Cosmic strings and domain walls in models with Goldstone and 
pseudo-Goldstone bosons'',
{\it Phys. Rev. Lett.} {\bf 48}, 1867-70 (1982)

\bibitem{Witten85} E. Witten, 
``Superconducting strings'',
{\it Nucl. Phys.} {\bf B249},  557-592 (1985)
 
\bibitem{DavisShellard89a} R.L. Davis, E.P.S. Shellard, 
``Cosmic vortons'',
{\it Nucl. Phys.} {\bf B323}, 209-2024 (1989)

\bibitem{Brandenbergetal96} R. Brandenberger, B. Carter, A.-C. Davis,
M. Trodden, ``Cosmic vortons and particle constraints'',
{\it Phys. Rev.} {\bf D54}, 6059-6071 (1996) 
[hep-ph/9605382]

\bibitem{ChamGib99} A. Chamblin, G. Gibbons,
``Supergravity on the brane'',
{\it Phys. Rev. Lett.} {\bf 84},  1090-1093 (2000)
[hep-th/9909130]
 
\bibitem{ChamEtal99}  A. Chamblin, S.W. Hawking, H.S. Real,
``Brane world black holes'',
{\it Phys. Rev.} {\bf D61}, 065007 (2000)
[hep-th/9909205]

\bibitem{BinEtal99} P. Binetruy, C. Deffayet, D. Langlois,
``Non-conventional cosmology from a brane universe'',
{\it Nucl. Phys.} {\bf B565},  269-287 (2000)
[hep-th/9905012]

\bibitem{ShirEtal00} T. Shiromizu, K. Maeda, M. Sasaki,
``The Einstein equations on the 3-brane world'',
{\it Phys. Rev.} {\bf D62}, 024012 (2000)
[gr-qc/9910076]

\bibitem{STW00} 
H. Stoica, H. Tye, and I. Wasserman,
``Cosmology in the Randall--Sundrum brane world scenario'',
 {\it Phys. Lett.} {\bf B482} (2000) 205 
[{\tt hep-th/0004126}].

\bibitem{GregEtal00} P. Bowcock, C. Charmousis, R. Gregory,
``General brane cosmologies and their global spacetime structure'',
{\it Class. Quant. Grav.} {\bf 17}, 4745 (2000)
[hep-th/0007177] 

\bibitem{Ms00} R. Maartens
``Cosmological dynamics on the brane'',
{\it Phys. Rev.} {\bf D62}, 084023 (2000)
[hep-th/0004166]

\bibitem{MLW00} D. Langlois, R. Maartens, D. Wands,
``Gravitational waves from inflation on the brane'',
{\it Phys. Lett.} {\bf B489}, 259-267 (2000)
[hep-th/0006007]

\bibitem{BatMen00} A. Mennim, R.A. Battye,
``Cosmological expansion on a dilatonic brane world''
[hep-th/0008192]

\bibitem{CarterBattye98} B. Carter, R. Battye,
``Non divergence of gravitational self interactions for Nambu-Goto 
strings'',  {\it Phys. Lett} {\bf B430},  49-53 (1998)
[hep-th/9803012]

\bibitem{GherghettaShaposhnikov00} T. Gherhgetta, M. Shaposhnikov,
``Localising gravity on a string-like defect in six dimensions'',
{\it Phys. Rev. Lett.} {\bf 85},  240-243 (2000)
[hep-th/0004014]

\bibitem{Carter92a} B. Carter, 
``Outer curvature and conformal geometry of an imbedding'',
{\it J. Geom. Phys.} {\bf 8}, 53-88 (1992).

\bibitem{Carter92b} B. Carter, 
``Basic brane theory'',
{\it J. Class. Quantum Grav.} {\bf 9}, 19-33 (1992)

\bibitem{Eisenhart26} L.P. Eisenhart, {\it Riemannian Geometry} 
(Princeton U.P., 1926, reprinted  1960)

\bibitem{Stachel80} J. Stachel, 
``Thickenning the string: the perfect string dust'',
{Phys. Rev.} {\bf D21}, 2171-81 (1980)

\bibitem{Schouten54} J.A. Schouten, {\it Ricci Calculus} (Springer, 
Heidelberg, 1954)

\bibitem{HawkingEllis73} S.W. Hawking, G.F.R. Ellis, {\it The Large Scale
Structure of Spacetime} (C.U.P., 1973)

\bibitem{CarterLanglois95} B. Carter, D. Langlois, 
``Kalb-Ramond coupled vortex fibration model for relativistic fluid dynamics'',
{\it Nucl. Phys.} {\bf B454},  402-424 (1995) [hep-th/9611082] 

\bibitem{CapovillaGuven95} Capovilla, R., and Guven, J.,
``Large deformations of relativistic membranes: a generalistion of
the Raychaudhuri equations'',
{\it Phys. Rev.} {\bf D52}, 1072 (1995) [gr-qc/9411061]

\bibitem{PenroseRindler84} R. Penrose, W. Rindler, 
{\it Spinors and Space-Time} (C.U.P., 1984)

\bibitem{BarsPope88} I. Bars, C.N. Pope,
``Anomalies in super p - branes'',
{\it Class. Quantum Grav.,} {\bf 5},  1157 - 1168 (1988)

\bibitem{Sikivie82} P. Sikivie, 
``Axions, domain walls, and the early universe'',
{\it Phys. Rev. Lett.} {\bf 48},  1156-59 (1982)

\bibitem{Shellard90}  E.P.S. Shellard,  
``Axion strings and domain walls'',
in {\it Formation and Evolution of Cosmic Strings,} 
ed. G. Gibbons, S. Hawking, T. Vachaspati,  107-115 
(Cambridge U.P., 1990)

\bibitem{Nambu77} Y. Nambu,
``String-like configurations in the Weinberg-Salam theory'', 
{\it Nucl. Phys.} {\bf B130},  505-515 (1977)

\bibitem{Manton83} N.S. Manton,
``Topology in the Weinberg-Salam theory'',
{\it Phys. Rev.} {\bf D28},  2019-26 (1983).

\bibitem{Copelandetal88a} E. Copeland, D. Haws, T.W.B. Kibble. D. Mitchel, N. Turok,
``Monopoles connected by strings'',
{\it Nucl. Phys.} {\bf B298},  458-492 (1988)

\bibitem{VachAch91} T. Vachaspati, A. Ach\'ucarro,
``Semilocal cosmic strings'', 
{\it Phys. Rev.} {\bf D44}, {\it pp} 3067-71 (1991)
 
\bibitem{VachBar92} T. Vachaspati, M. Barriola, 
``A new class of defects'',
{\it Phys. Rev. Lett.} {\bf 69},  1867-72 (1992)

\bibitem{MartinVilenkin96} X. Martin, A. Vilenkin,
``Gravitational background from hybrid topological defects''
{\it Phys. Rev. Lett.} {\bf 77}, 2879 (1996) [astro-ph/9606022].

\bibitem{DabholkarQuashnock90}  A. Dabholkar, J.M. Quashnock, 
``Pinning down the axion'',
{\it Nucl. Phys.} {\bf B333}, 815-832 (1990)
 
\bibitem{BattyeShellard95} R.A. Battye, E.P.S. Shellard, 
 ``String radiative backradiation", 
{\it Phys. Rev. Lett.} {\bf 75}, 4354-4357 (1995) [astro-ph/9408078].  

\bibitem{BattyeShellard96} R.A. Battye, E.P.S. Shellard,
``Radiative backreaction on global strings'',
{\it Phys. Rev.} {\bf D53}, 1811 (1996) [hep-ph/9508301]

\bibitem{Po86} A. Polyakov, 
``Fine structure on strings'',
{\it Nucl. Phys.} {\bf B268},  406-412 (1986)

\bibitem{MT88} K.I. Maeda, N. Turok, 
``Finite width corrections to the Nambu action for the Nielsen-Olesen string'',
{\it Phys. Lett.} {\bf B202}, 376-84 (1988) 

\bibitem{G88} R. Gregory,
``The effective action for a cosmic string'',
{\it Phys. Lett.} {\bf B206}, 199-204 (1988) 

\bibitem{G93} R. Gregory, 
``Effective actions for bosonic topological defects'',
{\it Phys. Rev.} {\bf D43}, 520-25 (1993)

\bibitem{Gal91} R. Gregory, D. Haws, D. Garfinkle, 
``Dynamics of domain walls and strings''.
{\it Phys. Rev.} {\bf D42}  343-345 (1991)

\bibitem{SM93} V. Silveira, M.D. Maia, 
``Topological defects and corections to the Nambu action'',
{\it Phys. Lett.} {\bf A174}, 280-288 (1993)

\bibitem{Bal94} C. Barrab\`es, B. Boisseau, M. Sakellariadou,
``Gravitational effects on domain walls with curvature corrections'', 
{\it Phys. Rev.} {\bf D49}, 2734-39 (1994)

\bibitem{CGr95} B. Carter, R. Gregory, 
``Curvature corrections to dynamics of domain walls'',
{\it Phys. Rev.} {\bf D51},  5839-46 (1995) [hep-th/9410095].

\bibitem{Let90} P.S. Letelier, 
``Nambu bubbles with curvature corrections'',
{\it Phys. Rev.} {\bf D41}, 1333-35 (1990)

\bibitem{HT90} D. H. Hartley, R.W. Tucker, in {\it Geometry of Low Dimensional
Manifolds, 1 }( {L.M.S. Lecture Note Series} {\bf 150}, 
ed. S. Donaldson, C. Thomas (C.U.P., 1990)

\bibitem{A92}  H. Arodz, A. Sitarz, P. Wegrzyn, {\it Act. Phys.
Polon. B}, {\bf 22}, 495 (1991); {\bf 23}, 53 (1992) 

\bibitem {BL92} B. Boisseau, P.S. Letelier,
``Cosmic strings with curvature corrections'', 
{\it Phys. Rev.} {\bf D46},  1721-29 (1992)

\bibitem {C94} B. Carter, 
``Equations of motion of a stiff geodynamic string or higher brane'',
{\it Class. Quantum Gravity} {\bf 11}, 2677-92 (1994)

\bibitem{CGu95} R. Capovilla, J. Guven, 
``Geomety of deformations of relativistic membranes'',
{\it Phys. Rev.} {\bf D51}, 6736 (1995) [gr-qc/9411060]. 

\bibitem{Carter94} B. Carter,
``Axionic vorticity variational formulation for relativistic perfect fluids'',
 {\it Class. Quantum Grav.} {\bf 11}, 2013-30 (1994)

\bibitem{VilenkinVachaspati87} A. Vilenkin, T. Vachaspati, 
``Radiation of Goldstone bosons from cosmic strings''	
{\it Phys. Rev.} {\bf D35}, 1138-40 (1987)
 
\bibitem{DavisShellard89} R.L. Davis, E.P.S. Shellard, 
``Global strings and superfluid vortices'',
{\it Phys. Rev. Lett.} {\bf 63}, 2021-24 (1989)

\bibitem{Sakellariadou91} M. Sakellariadou, 
``Radiation of Nambu-Goldstone bosons from infinitely long strings'',
{\it Phys. Rev.} {\bf D44}, 3767-73 (1991)

\bibitem{BenYaacov92} U. Ben-Ya'acov, 
``Unified dynamics of quantum vortices'',
{\it Nucl. Phys.} {\bf B382}, 597-615 (1992).                                       

\bibitem{Carter89} B. Carter, 
``Duality relation between charged elastic strings and superconducting 
cosmic strings'',
{\it Phys. Lett.} {\bf B224}, 61-66 (1989)

\bibitem{Larsen93} A.L. Larsen,
``A note on dispersive versus non-dispersive strings'', 
{\it Class. Quantum Grav.} {\bf 10}, L35-38 (1993)

\bibitem{Vilenkin91} A. Vilenkin, 
``Cosmic string dynamics with friction'',
{\it Phys. Rev.} {\bf D43}, 1060-62 (1991)

\bibitem{GarrigaSakellariadou93} J. Garriga, M. Sakellariadou, 
``Effects of friction on cosmic strings'',
{\it Phys. Rev.} {\bf 48}, 2502-15 (1993) [gr-qc/9307008]

\bibitem{Carteretal94} B. Carter, M. Sakellariadou, X. Martin, 
``Cosmological expansion and thermodynamic mechanisms in cosmic string dynamics'',
{\it Phys. Rev.} {\bf D50},  682-99 (1994)

\bibitem{KMPRS99} I.I. Kogan, S. Mouslopoulos, A. Papazoglou, G.C. Ross, 
J. Santiago, ``Three three - brane universe: new phenomenology for the new
millenium?'', {\it Nucl. Phys.} {\bf B584}, 313 -328 (2000)
[hep-ph/9912552] 

\bibitem{KMPR00} I.I. Kogan, S. Mouslopoulos, A. Papazoglou, G.G. Ross,
``Multi - brane worlds and modification of gravity at large scales''
[hep-th/0006030]

\bibitem{DDPV00} A.-C. Davis, S.C. Davis, W.B. Perkins, I.R. Vernon
``Brane world phenology and the Z$_2$ symmetry''
[hep-ph/0008132]

\bibitem{DerDol00} N. Deruelle, T. Dolezel,
``Brane versus shell cosmologies in Einstein and Einstein - Gauss - Bonnet
theories'', {\it Phys. Rev.} {\bf D62}, 103502 (2000)
[gr-qc/0004021]

\bibitem{Per00} W.B. Perkins, 
``Colliding bubble worlds''
[gr-qc/0010053]

\bibitem{Carter93} B. Carter, 
``Perturbation dynamics for membranes and strings governed by Dirac Goto Nambu 
action in curved space'',
{\it Phys. Rev.} {\bf D48}, 4835-38 (1993)

\bibitem{BattyeCarter95} R.A. Battye, B. Carter,
``Gravitational perturbations of relativistic membranes and strings'',
{\it Phys. Letters} {\bf B357}, 29-35 (1995)
[hep-ph/9508300]

\bibitem{BattyeCarter00} R.A. Battye, B. Carter,
``Second order Lagrangian and symplectic current for gravitationally perturbed 
Dirac-Goto-Nambu strings and branes'',
{\it  Class. Quant. Grav.} {\bf 17} 3325-3334 (2000)
[hep-th/9811075] 

\bibitem{Carter89a} B. Carter, 
``Stability and characteristic propagation speeds in superconducting cosmic 
and other string models'',
{\it Phys. Lett.} {\bf B228}, 466-470 (1989)

                                                        
\end{thebibliography}



\end{document}